\documentclass[manuscript]{aastex}

\shorttitle{Spectroscopy of SMC Red Giants III.}
\shortauthors{Parisi et al.}

\begin{document}

\title{Ca II Triplet Spectroscopy of Small Magellanic Cloud Red Giants.
III. Abundances and Velocities for a Sample of 14 Clusters}

\author{M.C. Parisi}
\affil{Observatorio Astron\'omico, Universidad Nacional de C\'ordoba}
\affil{Laprida 854, C\'ordoba, CP 5000, Argentina.}
\affil{Consejo Nacional de Investigaciones Cient\'ificas y T\'ecnicas}
\affil{Av. Rivadavia 1917, Buenos Aires, CP 1033, Argentina}
\email{celeste@oac.uncor.edu}

\author{D. Geisler}
\affil{Departamento de Astronom{\'\i}a, Universidad de Concepci\'on}
\affil{Casilla 160-C, Concepci\'on, Chile.}
\email{dgeisler@astro-udec.cl}

\author{J.J. Clari\'a}
\affil{Observatorio Astron\'omico, Universidad Nacional de C\'ordoba}
\affil{Laprida 854, C\'ordoba, CP 5000, Argentina.}
\affil{Consejo Nacional de Investigaciones Cient\'ificas y T\'ecnicas}
\affil{Av. Rivadavia 1917, Buenos Aires, CP 1033, Argentina}
\email{claria@oac.uncor.edu}

\author{S. Villanova}
\affil{Departamento de Astronom{\'\i}a, Universidad de Concepci\'on}
\affil{Casilla 160-C, Concepci\'on, Chile.}
\email{svillanova@astro-udec.cl}

\author{N. Marcionni}
\affil{Observatorio Astron\'omico, Universidad Nacional de C\'ordoba}
\affil{Laprida 854, C\'ordoba, CP 5000, Argentina.}
\email{nmarcionni@oac.uncor.edu}

\author{A. Sarajedini}
\affil{Department of Astronomy, University of Florida}
\affil{PO Box 112055, Gainesville, FL 32611, USA.}
\email{ata@astro.ufl.edu}

\and

\author{A.J. Grocholski}
\affil{Department of Physics and Astronomy, Louisiana State University}
\affil{202 Nicholson Hall, Tower Drive, Baton Rouge, LA 70803-4001, USA.}
\email{grocholski@phys.lsu.edu}

\begin{abstract}
We obtained spectra of red giants in 15 Small Magellanic Cloud (SMC) clusters in the region of the CaII lines with FORS2 on the Very Large Telescope (VLT). We determined the mean metallicity and radial velocity with mean errors of 0.05 dex and 2.6 km s$^{-1}$, 
respectively, from a mean of 6.5 members per cluster. One cluster (B113) was too young for a reliable metallicity determination and was excluded from the sample. We combined the sample studied here with 15 clusters previously studied by us using the same technique, 
and with 7 clusters whose metallicities determined by other authors are on a scale similar to ours. This compilation of 36 clusters is the largest SMC cluster sample currently available with accurate and homogeneously determined metallicities. We found a high probability that 
the metallicity distribution is bimodal, with potential peaks at -1.1 and -0.8 dex. Our data show no strong evidence of a metallicity gradient in the SMC clusters, somewhat at odds with recent evidence from CaT spectra of a large sample of field stars \citep{dob14}. This may be revealing possible differences in the chemical history 
of clusters and field stars. Our clusters show a significant dispersion 
of metallicities, whatever age is considered, which could be reflecting the lack of a unique AMR in this galaxy. None of the chemical evolution models currently 
available in the literature satisfactorily represents the global chemical enrichment processes of SMC clusters.

\end{abstract}


\section{Introduction}
The ages, abundances and kinematics of star clusters are prime
indicators of a galaxy's chemical evolution and star formation history (SFH).
This is particularly true for the Small Magellanic Cloud (SMC), which is close enough to provide a wealth
of detail in studies including even its oldest stellar populations and also hosts a huge star cluster ensemble.
These star clusters also have importance to astronomy beyond
the bounds of the SMC.  Because of their richness and location in areas of the age-metallicity plane not
covered by Milky Way clusters, they have become vital testbeds
for theoretical models of stellar evolution at young to intermediate age and low metallicity (e.g.,
\citealt{fer95,whi03}).  SMC clusters
have also been used as empirical templates for interpreting
the unresolved spectra and colors of very distant galaxies,
including post-starbursts and other pathological cases
(e.g., \citealt{bea02,bic86}).  The interaction
history of the Magellanic Clouds (MCs) with the Galaxy is a matter of current controversy \citep{kal13}, and clusters
can serve as important keystones to help pin down the epoch(s) of increased cluster and field star formation
due, e.g., to a close galactic encounter.

A major objective is to measure many clusters spanning as wide a range of age, abundance and location as possible in order
to maximize our leverage on the chemical evolution as traced by the age-metallicity relation (AMR), metallicity gradient, kinematics and  any variation of SFH with location and/or environment.
It is also paramount  to definitively test for the existence
of any bursts of cluster formation (e.g., \citealt{ric00}). 

The many
free parameters inherent in any realistic chemical evolution model, e.g., that allows radial variations,  demands as large a cluster sample as possible.  In addition, one requires a homogeneous technique
with sufficient precision and accuracy in both age and metallicity.
The brightest common stars in clusters older than
$\approx$1~Gyr are red giants.  Therefore, they are the natural
targets for precision measurements of abundances and velocities, especially in extragalactic clusters.
The most efficient way to build up a
large sample of red giant metallicity and velocity measurements is by using
the near-infrared Ca~II triplet (CaT) at $\lambda$ $\approx$ 8500 \AA, which requires only very moderate
resolution spectra as the lines are very strong and very abundance sensitive, and this is near the peak in the flux of red giants.
Multi-object spectrographs add an extra dimension of efficiency.
Many authors have confirmed the accuracy and repeatability
of CaT abundance measurements  in combination
with broad-band photometry, both optical \citep{ard91} and IR \citep{mau14}, and shown its insensitivity to age \citep[hereafter C04]{col04} and sensitivity to metallicity.

Our group has carried out a number of studies of MC clusters using the powerful CaT technique
with FORS2 on the VLT. Our first   study \citep[hereafter G06]{gro06} yielded excellent
data for 28 LMC clusters.
We found a very tight metallicity
distribution (MD) for intermediate-age clusters, no metallicity gradient and
confirmed that the clusters rotate with the disk. We followed this up with an
initial study of SMC clusters \citep[hereafter P09]{par09}. We obtained spectra for 102 stars
associated with 16 SMC clusters. Based on the color-magnitude diagrams (CMDs) from
the preimages, spatial distribution, metallicity and velocity analysis, we were
able to separate cluster from field stars with very high probability. We
determined mean cluster velocities to 2.7 kms$^{-1}$ and metallicities to 0.05 dex
(random error), from a mean of 6.4 members per cluster.
We combined our clusters with those observed by \citet[hereafter DH98]{dch98}, the only previously published CaT
study for SMC clusters, and with clusters studied by \citet{kay06}, whose study has not been published but whose metallicity values are reported in \citet{gla08b}.
 We found a suggestion
of bimodality in the MD and no evidence for a metallicity gradient.
The AMR showed
good overall agreement with the model of \citet[hereafter PT98]{pag98} which assumes a burst of star formation at 4 Gyr,
except for two clusters around 10 Gyr which are more metal-rich than the prediction
and our 4 youngest clusters, which all lie to lower metallicities than predicted.
The simple closed box model of DH98 yields a much poorer fit.
The two ``anomalous'' older clusters are L1 and K3, observed by DH98, who were limited
by the technology at the time to single star spectra and could only observe with a 4 m telescope a total of 4 stars per cluster.
We also examined the kinematics and found no obvious signs of rotation. Simultaneously,
we obtained similar quality radial velocity and metallicity data for $\sim 300$ surrounding field giants.
The results for these stars were presented in \citet[- P10]{par10}.

However, as shown in P09 and \citet[hereafter P14]{par14}, it is clearly necessary to increase the number of SMC star clusters homogeneously studied for a better understanding of  the evolution and chemical enrichment processes in this galaxy. Up to this moment, there are only two SMC clusters whose metallicities have been determined
from high dispersion spectroscopy and only about 20 with metallicities derived from CaT spectroscopy (P09, DH98).
The remaining metal abundance determinations are based on less precise photometric and integrated spectroscopic techniques.
Here we present similar excellent CaT data for a new sample of SMC star clusters using the same telescope and
instrument as P09 in order to improve our knowledge of the AMR. We also reobserve the problematic 
clusters L1 and K3 originally observed by DH98 mentioned above in order to help pin down the chemical evolution during this important early phase.  This data yields comparable high precision mean metallicity
and radial velocity values per cluster as derived in P09. Together, this comprises the first dataset with sufficient
precision and accuracy  in metallicities and enough clusters to really test chemical evolution models.
Secondly, we can better probe previous hints that the metallicity and/or
 age distributions of SMC clusters exhibit
bimodality or similar fine structure, as well as spatial variation in metallicity.

In Section 2 we present our cluster and target selection and in Section 3 the spectroscopic observations are described.
Sections 4 and 5 give details about the measurement of radial velocities, equivalent widths and target metallicities, while Section 6 describes the procedure
to separate cluster stars from those belonging to the surrounding field.
In Section 7 we compare our metallicity determinations with the values found by previous work,
in Sections 8 we analyze the metallicity and finally, in Section 9 we summarize our results.

\section{Cluster and target selection}

In order to increase the number of SMC clusters homogeneously studied by P09, we selected an additional sample of 15 star clusters spread out over a wide region
of the SMC. Our clusters are scattered across the main body of the SMC, in environments ranging
from the dense central bar to near the tidal radius.
In P09 we found a difference between
the predictions of the PT98 model and the observations in the 9-10 Gyr age range (L1 and K3, which were observed by DH98) and
here we remeasure the abundances of these two
clusters to confirm their place in the age-metallicity plane.
We also observe a number of old cluster candidates in order to populate
this region as much as possible. We have selected all of the richest clusters with ages in the 5 - 10 Gyr range from
the work of \citet{rz05}.
There is also a significant discrepancy between the PT98 model and the P09 data for clusters with
ages $\sim1$ Gyr.
We are confident that this latter is not the result of significant age effects on the CaT
method (C04, \citealt{car08}), and therefore also include  more $\sim$ 1 Gyr old
clusters to investigate the extent of any discrepancies.
If confirmed, this would imply a very recent
infall of mostly primordial gas into the SMC, which has
implications for the chemical evolution of not just the SMC but also
potentially the LMC and our Galaxy as well.

\begin{figure}
\plotone{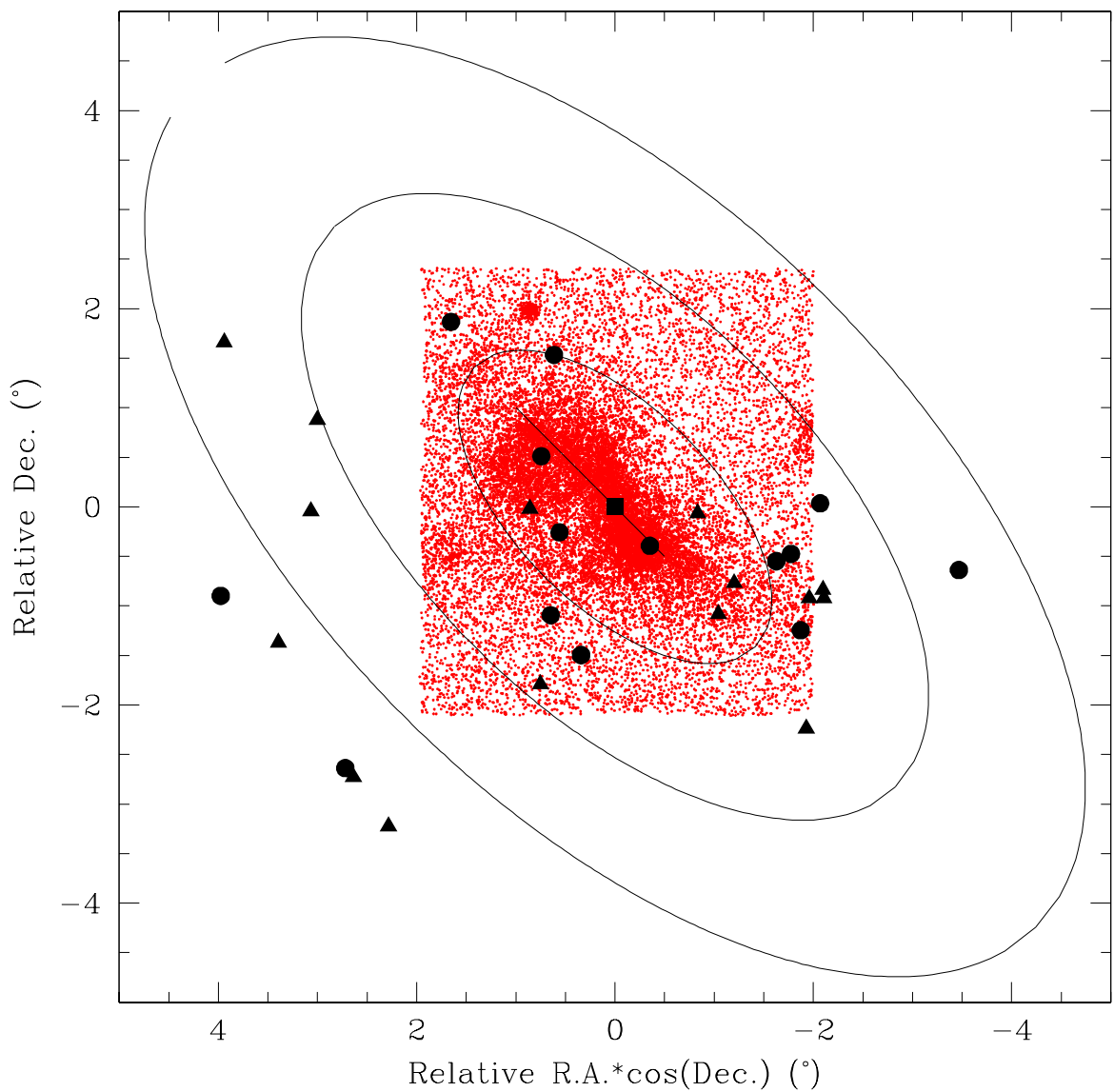}
\caption{ Circles indicate the position of our target clusters in relation to the SMC optical center (square, $\alpha_{J2000}$ = 00$^h$ 52$^m$ 45$^s$ and 
$\delta_{J2000}$ = 72$^{\circ}$ 49' 43", \citealt{cro01}) and the SMC bar (line). Triangles
represent the position of clusters studied in P09. The ellipses have
semi-major axis of 2, 4 and 6 degrees. A SMC map from \citet{zar02} has been superposed (Small Magellanic Cloud Photometric Survey for stars with V $<$ 16.)}
\label{f:elipse}
\end{figure}

In Table \ref{t:sample} we present the clusters selected for observation. Included are the identification of these clusters, their equatorial
coordinates, the semi-major axis $a$ \citep{pia07a,pia07b} and the age adopted as well by the respective bibliographic reference.  
Considering that the  reliability of \citet{gla10}'s age determination for cluster K9 and B113 (0.5 and 0.6 Gyr, respectively, the only age values available in the literature) 
is probably 
lower than that of the other cluster sample, we adopt the preliminary age (1.09 $\pm$ 0.15 Gyr for K9 and 0.53 $\pm$ 0.07 for B113) derived in our work currently in progress \citep{par15}. 
These values were obtained from the $\delta$V parameter measured on the cluster CMD and from the calibration of \citet{cc94}, using a procedure similar to the one 
described in P14. B113 turned out to be too young to reliably apply the CaT technique, so we decided to discard this cluster from our sample.
For the sake of consistency with P09,
we have adopted the elliptical system introduced by \citet{pia05}, in which the corresponding semi-major axis $a$ is used instead of the projected distance to the
galaxy center. Although the $a$ values do not consider projection effects, they represent the SMC shape better than a circular system.
When selecting the clusters,
we covered as large a spatial extension within the SMC as possible so as to examine, among other things, possible metallicity gradients in this galaxy.
Figure \ref{f:elipse} shows how the selected clusters (circles) are distributed with respect to the SMC bar and center. Clusters studied in P09 (triangles) have 
also been included in this figure to allow comparison between the two samples. \\

Spectroscopic targets in each cluster were selected from CMDs built from aperture photometry of preimages in the $V$ and $I$ bands, in the same way as described
in detail in P09.

\section{Spectroscopic observations}

As part of the ESO programs 082.B-0505 and 384.B-0687, spectra of more than $\sim$ 450 stars were obtained in service mode.
We used the FORS2 instrument in mask exchange unit (MXU) mode on the Very
Large Telescope (VLT), located at Paranal (Chile). Our instrumental setup was identical to that used in G06 and P09,
wherein more detailed description can be found, in order to ensure homogeneity.

The spectrum of each star was obtained with exposures of 680 and 635 seconds during the first and second observing runs, respectively. We used slits 1'' wide
and between 2'' and 12'' long. The seeing was less than 1'' in all cases. Pixels were binned 2x2, resulting in a scale of  0.25''/px. The spectra cover a
range of $\sim$ 1500 \AA \space in the region of the CaT ($\sim$ 8500 \AA) and have a dispersion of 0.85 \AA /px. Most cases have S/N ratio between $\sim$ 20 and $\sim$ 80
pixel$^{-1}$, with only a few stars having S/N $\sim$ 15 pixel$^{-1}$. Calibration exposures,
bias frames and flat-fields were also taken by the VLT staff. We followed the image processing detailed in G06 and P09, using a variety of tasks from the IRAF package.

\section{Radial velocity and equivalent width measurements}

Our reduction and analysis techniques are identical to those applied in G06 and P09.
The program used to measure the equivalent widths (EWs) of the CaT lines uses as input the radial velocity (RV) of the target stars to  derive the CaT line centers. These RVs  allow us help discriminate probable cluster members
from stars belonging to their surrounding stellar fields (see below).\\

To measure the RVs of our program stars, we performed cross-correlations between their spectra and those of 32 bright template giants observed in Milky Way clusters using the IRAF task {\it fxcor} \citep{ton79}. This task also transforms observed RVs into heliocentric RVs. We used the template stars of C04
who observed these stars with a setup very similar to ours. The average of each cross-correlation result was adopted as the heliocentric
RV of each target. Our heliocentric RVs have a typical standard deviation of $\sim$ 6 km s$^{-1}$.\\

It is well known that errors in centering the image in the spectrograph slit can lead to inaccuracies in determining RVs (e.g., \citealt{irw02}).
In order to correct this effect,
we measured the offset $\Delta x$ between each star's centroid and the corresponding slit center by inspecting the through-slit image taken immediately before
the spectroscopic observation, according to the procedures described by C04 and G06. We calculated the velocity correction $\Delta v$ according to equation (1)
from P09. We estimated a measurement precision of 0.14 pixels. Considering our spectral resolution of 30 km s$^{-1}$ px$^{-1}$, the typical error introduced
in the RV, by this effect, turns out to be 4.2 km s$^{-1}$. We added this error in quadrature with the one resulting  from the cross-correlation. This yields a total of 7.3
 km s$^{-1}$, which has been adopted as the typical RV uncertainty of an individual star.\\

To measure EWs, we have used a program whose details can be seen in C04. We adopted the rest wavelength of the CaT lines from
\citet{arz88}, along with the corresponding continuum bandpasses (see G06's Table 2). The ``pseudo-continuum'' for each CaT line was determined by a linear
fit to the mean value in each pair of continuum windows. The ``pseudo-equivalent width''  was calculated by fitting a function to each CaT line with respect
to the pseudo-continuum. For spectra with S/N $>$ 20, we fitted to each CaT line a Gaussian $+$ Lorentzian function in order not to underestimate the strength of the wings of the
line profile (\citealt{ru97a,ru97b}, C04, P09). Then we calculated the metallicity index, $\Sigma W$, defined as the sum of the EWs of the three CaT lines.
For spectra with S/N $<$ 20, however, we followed the procedure described in P09, i.e., we fitted only a Gaussian to each CaT line and we then corrected $\Sigma W$
 according to equation (2) from P09.

\section{Metallicities}

The procedure to determine the metallicity of a giant star from CaT lines and the required calibrations are described in detail in G06 and P09. In summary, since our spectra are of high enough quality for all three CaT lines to be well measured,
we calculated $\Sigma W$ using the following expression:

\begin{equation}
\Sigma W = EW_{8498} + EW_{8542} + EW_{8662},
\end{equation}

\noindent in which equal weight was assigned to all three lines. Then, we defined the reduced equivalent width, $W'$, to remove the effects of surface gravity
and temperature on $\Sigma W$ via its luminosity dependence:

\begin{equation}
W' = \Sigma W + \beta (v-v_{HB}).
\end{equation}

The difference between the visual magnitude ($v$) of the star and the cluster's horizontal Branch/red clump ($v_{HB}$) also removes any dependence on cluster
distance and interstellar reddening. We measured this magnitude difference using aperture photometry performed on the {\it pre-images}, which were uncalibrated, hence the use of lower-case letters to
denote the photometry.
Finally, the metallicities of the whole cluster sample were derived from the following relationship:

\begin{equation}
[Fe/H] = b + a \times W'.
\end{equation}

\noindent We used the values of $\beta = 0.73 \pm 0.04$, $b = -2.966 \pm 0.032$ and $a = 0.362 \pm 0.014$, taken from C04, who used an
instrumental setup similar to ours to derive them. This is the same calibration used in P09, ensuring our metallicities are homogeneously determined. 
The C04 calibration is nominally valid in the age range of 2.5 Gyr $\leq$ age $\leq$ 13 Gyr and in the metallicity range
-2.0 $\leq$ [Fe/H] $\leq$ -0.2. \\

Following this procedure, we derived individual metallicities of the observed red giant stars with errors ranging from 0.09 to 0.32 dex, with a mean of 0.16 dex.

\section{Cluster membership}

To discriminate cluster members from non-member stars belonging to the surrounding field as much as possible, we followed the procedure described in G06 and P09. Using the
coordinates from the aperture photometry, we determined the center of each cluster by building projected histograms in the $x$ and $y$ directions. We then fitted
Gaussians to these histograms (using the gnuplot program) and adopted the center of these Gaussians as the cluster center. We then obtained the cluster radial profile based on star counts
carried out over the entire area around each cluster.\\

\begin{figure}
\plotone{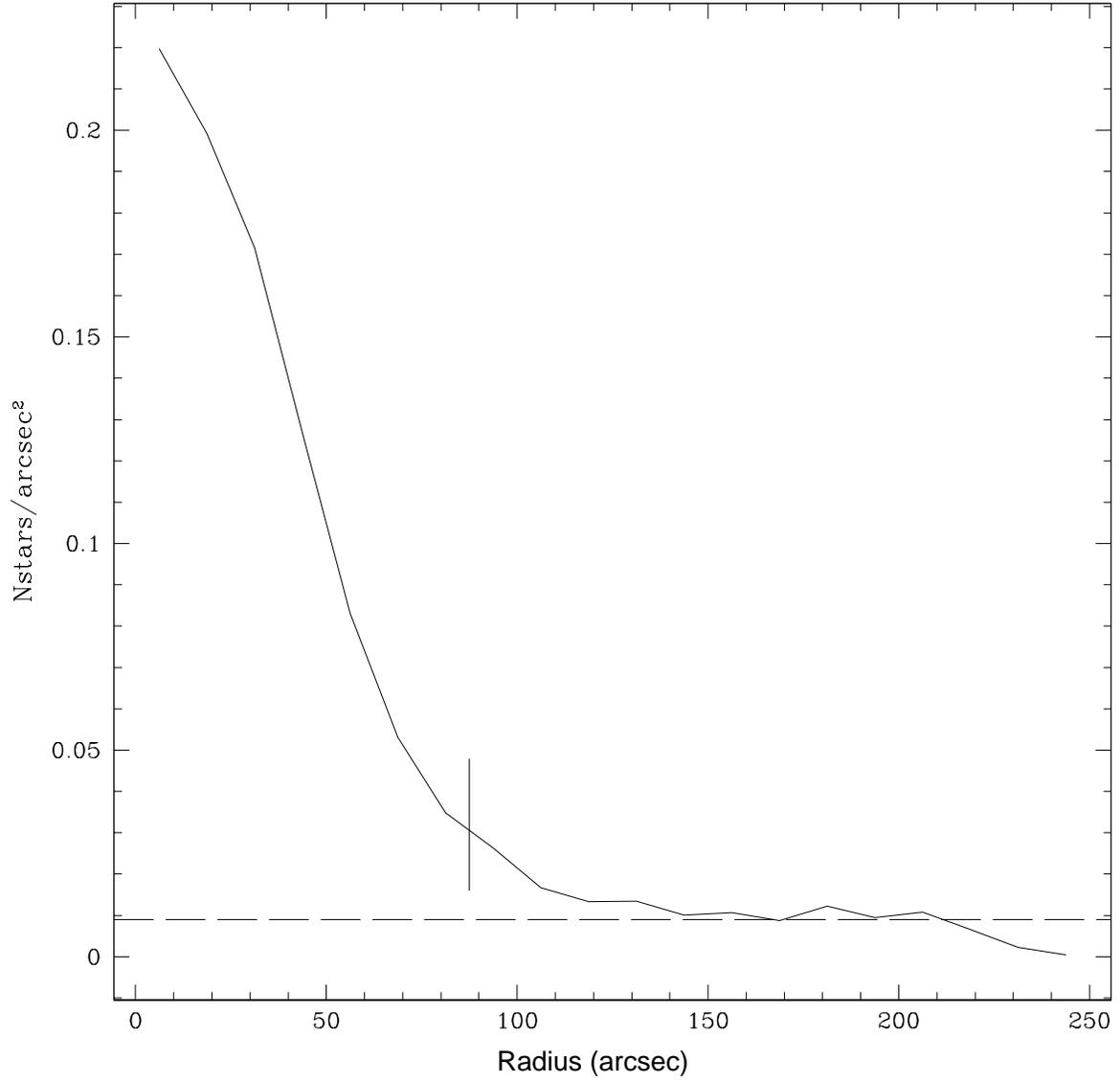}
\caption{Radial stellar density profile of cluster K\,3. The radius adopted for the
cluster is indicated by the vertical line. The background level is marked by the
dashed horizontal line.}
\label{f:prf}
\end{figure}

As an example to illustrate the process employed for all clusters, we show in Figure \ref{f:prf}  the radial profile obtained for cluster K\,3. The vertical line on
the profile represents the adopted cluster radius, which will be used in the subsequent analysis to evaluate cluster membership. Note that the adopted
cluster radius can differ from the more typical definition, in which the radius is the distance from the center to the point where the stellar density
profile reaches that of the background \citep{pia07c}. In our analysis, we adopted in most cases smaller radii than those resulting from this
definition in order to maximize the probability of cluster membership.\\

Once the cluster radii were determined, we built instrumental CMDs for each cluster using aperture photometry derived from our $v –$ and $i-$band {\it pre-images}.
These CMDs were constructed using only the stars located within the apparent radius. Then, we measured $v_{HB}$ as the median value of all stars inside of a box
that is 0.7 mag in $v$ and 0.3 mag in $v-i$ and centered on the red clump (RC) by eye. We preferred to use the median value instead of the mean value for the
reasons stated in P09. Errors in $v_{HB}$ are taken as the standard errors of the median.\\

\begin{figure}
\plotone{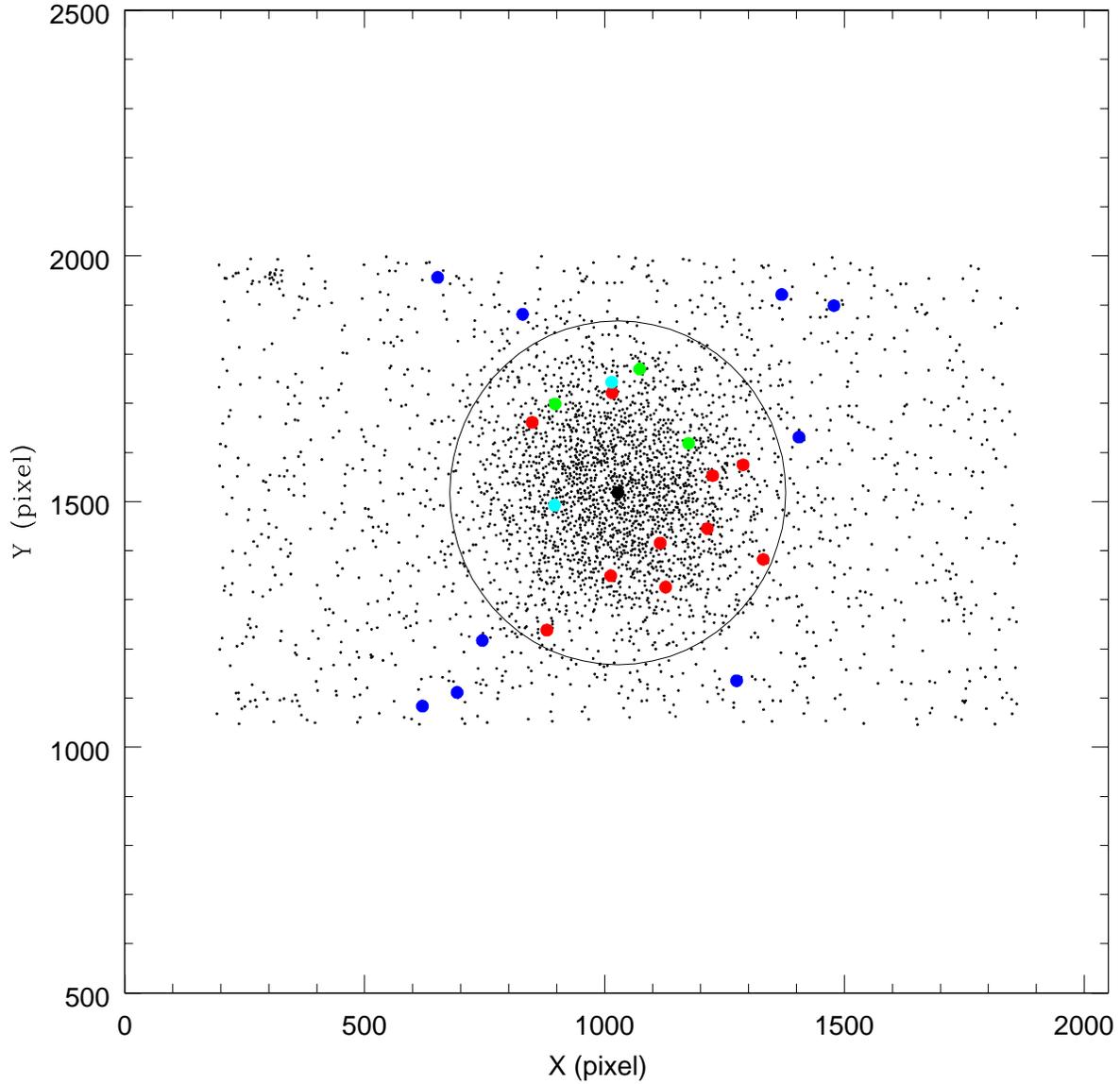}
\caption{Schematic finding chart of cluster K\,3. Our spectroscopic target stars
are represented by the large filled circles and the adopted cluster radius is
represented by a large circle.  Blue circles indicate non-members outside the
cluster radius. Teal and green circles represent non-members eliminated due
to discrepant RV or metallicity, respectively. Red circles indicate final cluster members. }
\label{f:chart}
\end{figure}

As the first step to evaluate cluster membership, we considered as non-members those stars located outside the adopted cluster radius. As an example,
we plot in Figure \ref{f:chart} the $xy$ positions of all stars photometrically observed in and around K\,3. Our spectroscopic targets are represented by the large filled
symbols, and the adopted cluster radius is indicated by the large circle. The target stars in blue are considered non-members due to their location
outside the adopted cluster radius.\\

\begin{figure}
\plotone{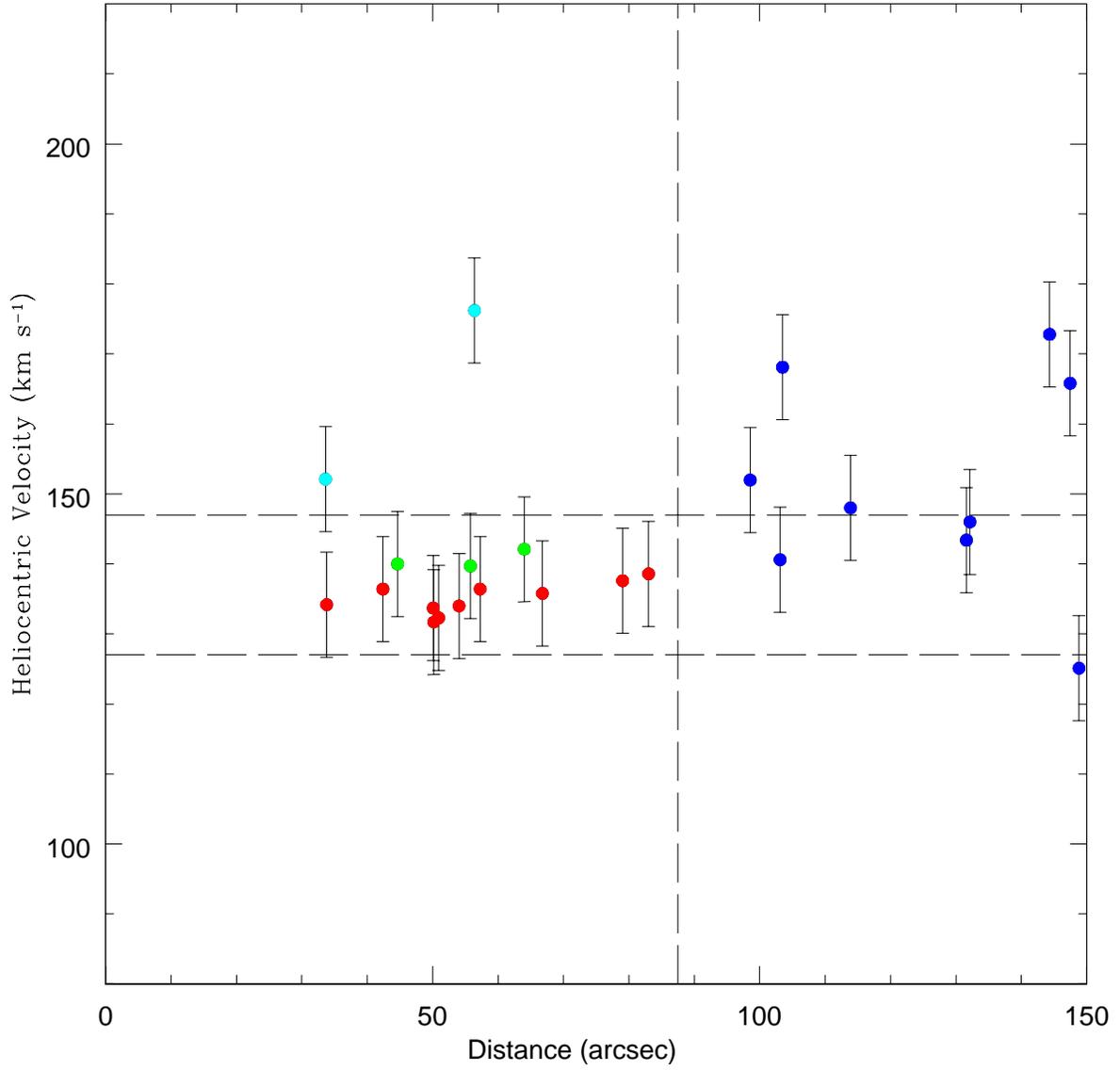}
\caption{Radial velocity vs. distance from the cluster center for K\,3 targets.
Horizontal lines represent our velocity error cut and the vertical line the
adopted cluster radius. Color code is the same as in Figure \ref{f:chart}.}
\label{f:vels}
\end{figure}

The second and third steps to discriminate cluster members from non-members was to analyze the behavior of the RVs and metallicities as a function of distance
from the cluster center. Figure \ref{f:vels}  shows how the RVs of the observed stars in K\,3 vary as a function of clustercentric distance. We have
adopted an intrinsic cluster velocity dispersion of 5 km s$^{-1}$ \citep{pry93}, which added in quadrature with our adopted RV error (7.3 km s$^{-1}$),
yields an expected dispersion of $\sim$ 9 km s$^{-1}$. We have rounded this up and adopted 10 km s$^{-1}$ in our analysis. Horizontal lines in Figure \ref{f:vels} show our cuts 
in RV = $\pm$ 10 km s$^{-1}$, which represents the expected RV dispersion within the cluster. These RV cuts were adopted taking into account the fact that
the members of a cluster should have a velocity dispersion lower than that of the field stars. The vertical line in Figure \ref{f:vels} again shows the adopted cluster radius.\\

\begin{figure}
\plotone{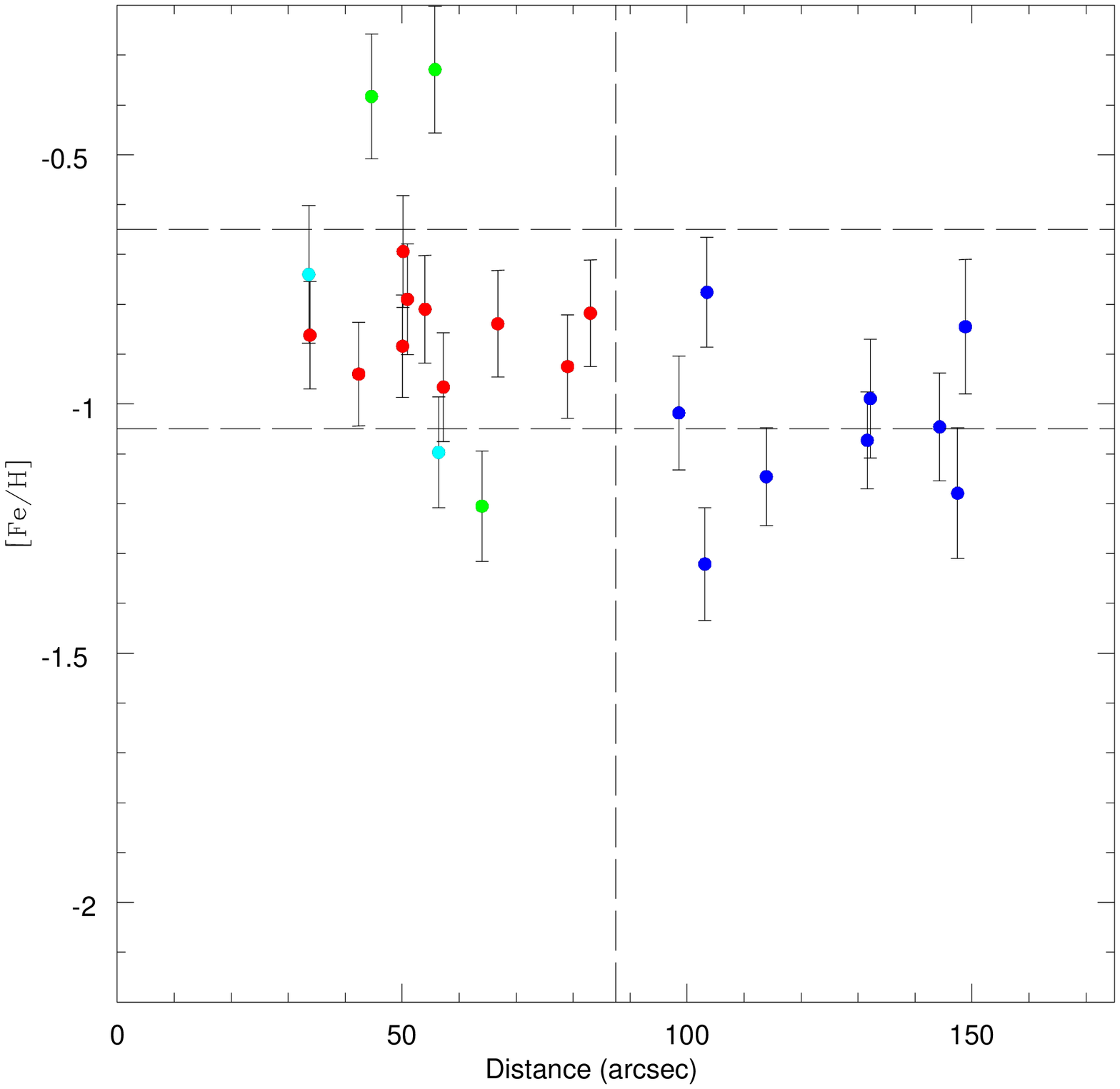}
\caption{Metallicity vs. distance from the cluster center for K\,3 targets. Horizontal lines represent the [Fe/H] error cut and the vertical line the adopted cluster radius. Color code is the same as in Figure \ref{f:chart}.}
\label{f:mets}
\end{figure}

\begin{figure}
\plotone{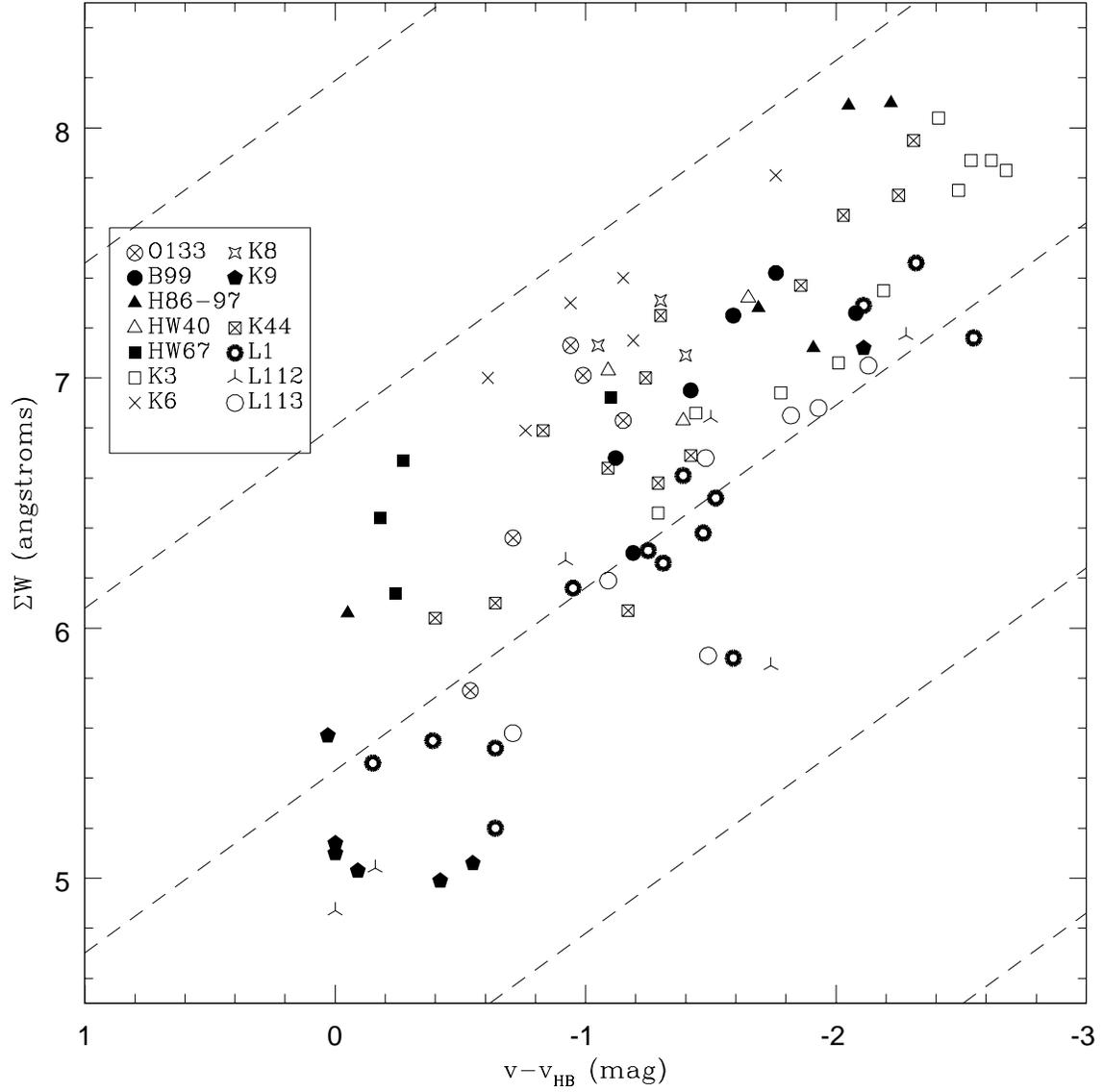}
\caption{ $\Sigma W$ vs. $v-v_{HB}$ for members identified in each cluster (represented by different symbols).  Isometallicity 
lines of 0, $-$0.5, $-$1, $-$1.5 and $-$2  (from top to bottom) are included.}
\label{f:isoab}
\end{figure}

Figure \ref{f:mets}  shows how the metallicities of the observed stars vary as a function of the distance from the center of K\,3. Since we have estimated a typical metallicity 
error of 0.16 dex for an individual star, we adopted an [Fe/H] error cut of $\pm$ 0.20 dex, represented by horizontal lines in Figure \ref{f:mets}. As in Figure \ref{f:vels}, the vertical line
in Figure  \ref{f:mets}  represents the cluster radius. Following the G06 and P09 color code, we have adopted in Figures  \ref{f:vels}  and  \ref{f:mets}  blue symbols to represent non-members lying outside
the cluster radius, teal and green symbols for non-members we eliminated for having RV and metallicity discrepancy, respectively, and red symbols for stars
that have passed all three cuts and are therefore considered cluster members. This procedure has been followed for each cluster in our sample. 
Figure \ref{f:isoab} shows $\Sigma W$ vs. $v-v_{HB}$ for our cluster sample.\\

Table \ref{t:giants} also shows the identification of the star, heliocentric RV and its error in columns (1), (2) and (3), $v-v_{HB}$ in column (4), $\Sigma W$ and 
its error in columns (5) and (6) and metallicity and its error in columns (7) and (8). \\

Finally, using these member stars, we calculated the mean cluster RV and the mean cluster metallicity and their standard error of the mean. The final results
are presented in Table \ref{t:results} where we successively list: cluster ID, the number $n$ of stars considered to be members, the mean heliocentric RV with its error and the
mean metallicity followed by its error. Errors in RV and metallicity correspond to the standard error of the mean. 

\section{Comparison to previous metallicity determinations}

Before analyzing our metallicity results, it is important to see how they compare with any previous determinations by other authors. The current
CaT metallicities for B\,99, H\,88-97, K\,8, K\,9 and OGLE\,133 appear to be the first metallicity determinations made for these clusters. Only three (K\,3,
L\,1 and L\,113) out of the remaining nine clusters of our sample have spectroscopic metallicities determined by DH98 from CaT, whereas the metallicity for HW\,40,
K\,3, K\,6 and L\,113 has been estimated from integrated spectroscopy \citep{pia05, dia10,dia14}.
Other metallicity determinations reported in the literature, for the previously studied clusters, are based on photometric techniques, mostly on  the Washington photometric
system \citep{pia11a,pia11b,pia11c,mig98,pia01,pia07b}.

\begin{figure}
\plotone{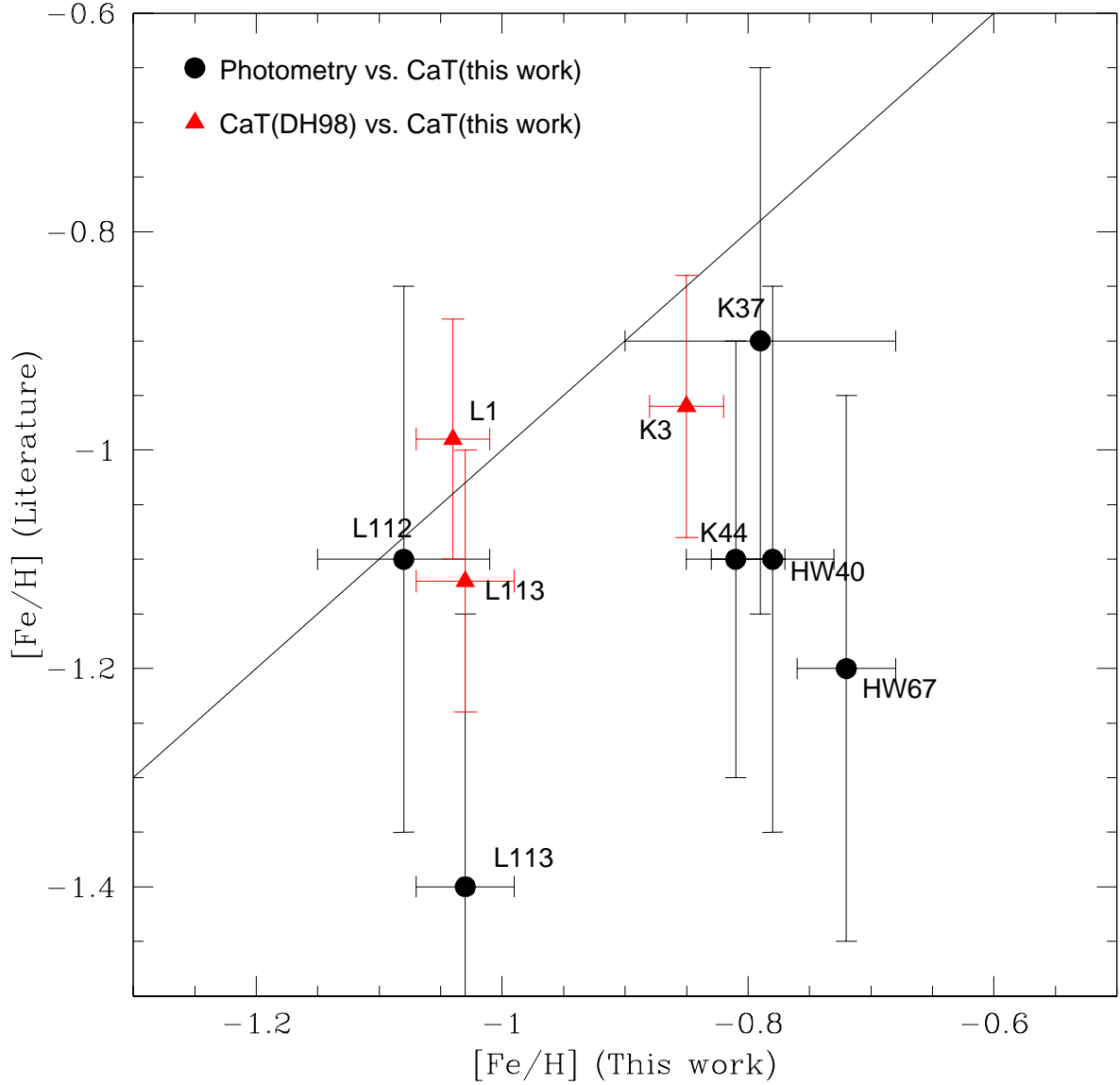}
\caption{Comparison of our spectroscopic mean cluster metallicities and those derived by other authors from Washington photometry (filled circles)
and by DH98 from CaT (triangles). The line shows one-to-one correspondence. }
\label{f:met_comp}
\end{figure}

\begin{figure}
\plotone{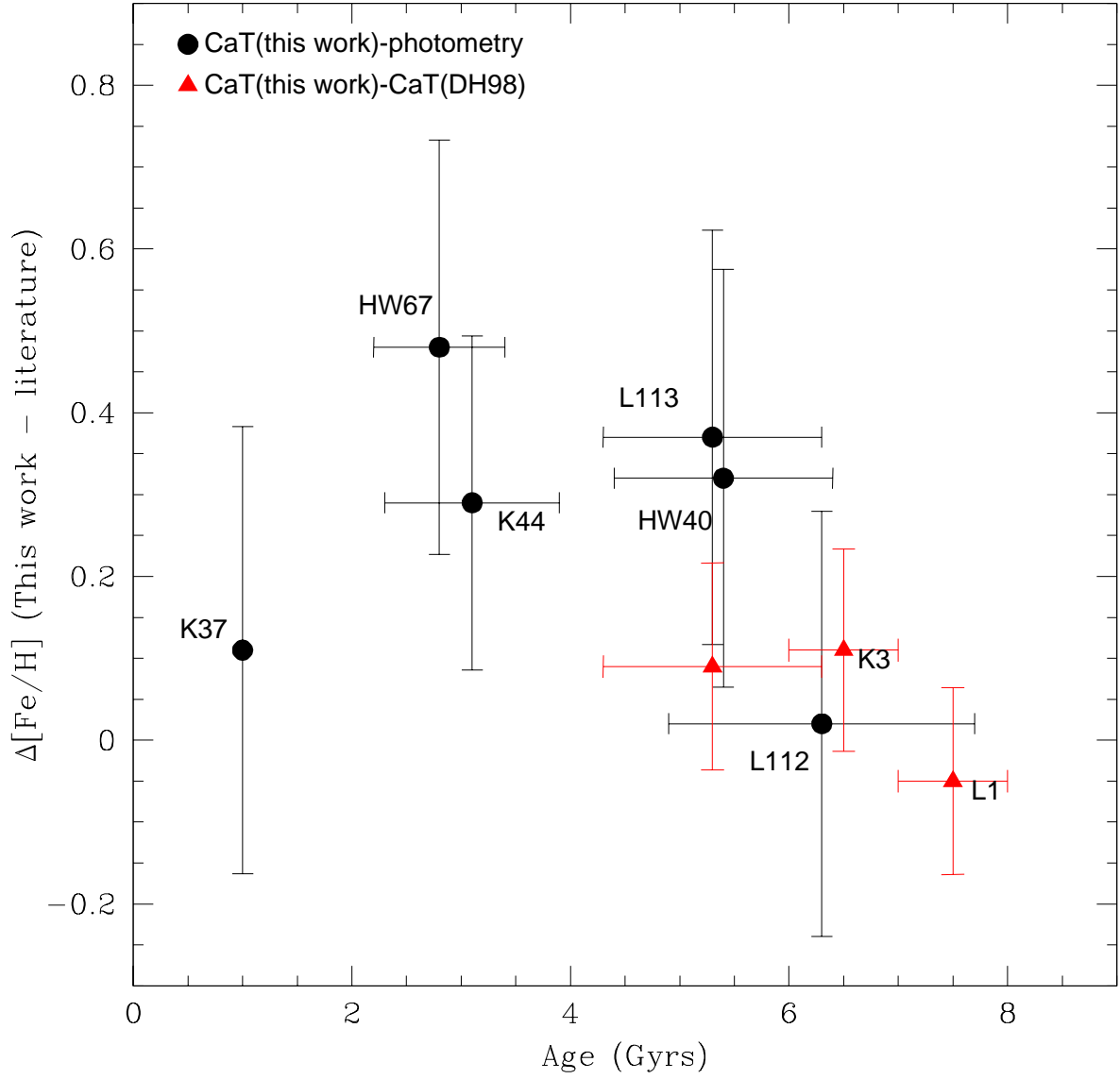}
\caption{Difference between our CaT metallicities and those derived by other authors as a function of age. Symbols are the same as in Figure \ref{f:met_comp}.}
\label{f:met_comp_dif}
\end{figure}

In Figure \ref{f:met_comp}, we have plotted the metallicities available in the literature as a function of those derived from our present work for the clusters in common.
The solid line indicates one-to-one correspondence. In addition, Figure \ref{f:met_comp_dif} shows the difference between both metallicity values as a function of cluster age. 
In Figures \ref{f:met_comp} and \ref{f:met_comp_dif}, filled circles stand for clusters whose metallicities haven been previously determined by other authors from Washington photometry. Triangles represent the three clusters for which DH98 derived CaT metallicities. Our CaT metallicities are
in excellent agreement with those found by DH98, with the difference between them generally smaller than 0.1 dex, certainly within the respective errors, and with no systematic offset. This gives us added confidence that our metallicities are on solid footing.
On the other hand, the mean difference (in absolute value)
between our metallicities and those derived from Washington photometry is 0.26 $\pm$ 0.17 dex, with our values being more metal-rich.
In P09, we also compared our CaT-based metallicities with those based on Washington photometry for 11 clusters and found no systematic difference. Here all 6 Washington metallicities are lower than our
values. However, the differences are not very significant given the much larger photometric metallicity
error bars. Because of the age-metallicity degeneracy, a significant variation of metallicity can affect cluster ages when they are determined by theoretical isochrones, 
which may, in turn, affect the conclusions drawn about the AMR. We plan to determine the ages of the clusters studied in this paper following the same procedures as in P14, in a subsequent paper.

Regarding metallicity determinations from integrated spectroscopy, we find a reasonable agreement with our  values for clusters HW\,40 and K\,6. However, this agreement
is substantially poorer for clusters L\,113 and K\,3. There are several previous metallicity determinations from integrated spectroscopy for these two clusters. From those
determinations, the reported values of [Fe/H] are -1.2, -1.8 \citep{pia05,dia10} for K\,3 and -1.4, -2.6 \citep{dia10} for L\,113. The minimum difference between both spectroscopic metallicity determinations (integrated and CaT) is 0.35 dex; it is much larger, however, if the most metal-poor  values in the literature are adopted.
For the integrated spectra technique, it is harder to assess the significance of the offset with our values given the lack of errors for some of the clusters.
We note that the CaT technique is generally believed to be more robust, less sensitive to age and observing technique and model-independent and so should be given higher weight.

Figure  \ref{f:met_comp_dif} shows there is no significant trend for the offset in metallicity with cluster age.

\section{Metallicity analysis}

In order to best analyze the SMC chemical properties, it is  optimal not only to have available a cluster sample larger than the one we investigate here but also
to maintain homogeneity as much as possible. For this  reason, we added to the present sample other clusters having well-determined metallicities on a scale judged to be the same as, or similar to, our's. As mentioned, in P09 the metallicity of 15 SMC clusters was determined following exactly the same procedure as in this study. It is therefore  appropriate to add them to the present sample. The ages we adopt for these 15 clusters are those determined by P14 from the $\delta V$ parameter using
the \citet{cc94} calibration. In addition, our sample has been complemented by the L\,11, NGC\,121 and NGC\,339  metallicities from DH98 and ages based on deep HST data
\citep{gla08a,gla08b}. The corresponding metallicity (converted to the scale of \citealt{cag97}) and age values reported by \citet{gla08b} for NGC\,416, L\,38 and NGC\,419 were also included. In both studies, metallicities have been
inferred from the CaT technique. Thus, we have a sample consisting of 35 clusters homogeneously studied with accurately determined metallicities. It is important to remark 
that this is at present the largest SMC cluster sample with well-determined metallicities on a homogeneous scale. As part of our sample, we also included NGC\,330, 
whose metallicity was obtained from high dispersion spectroscopy (-0.94 $\pm$ 0.02, \citealt{gon99}). We note that \citet{hil99} also studied NGC\,330 with high dispersion spectroscopy, finding a slightly higher value (-0.82 $\pm$ 0.11 dex). Both values are in agreement within the errors but the
\citet{gon99} is more precise so we adopt their value, as well as for consistency with our previous
works (P09, P14).


\subsection{Metallicity distribution}

\begin{figure}
\plotone{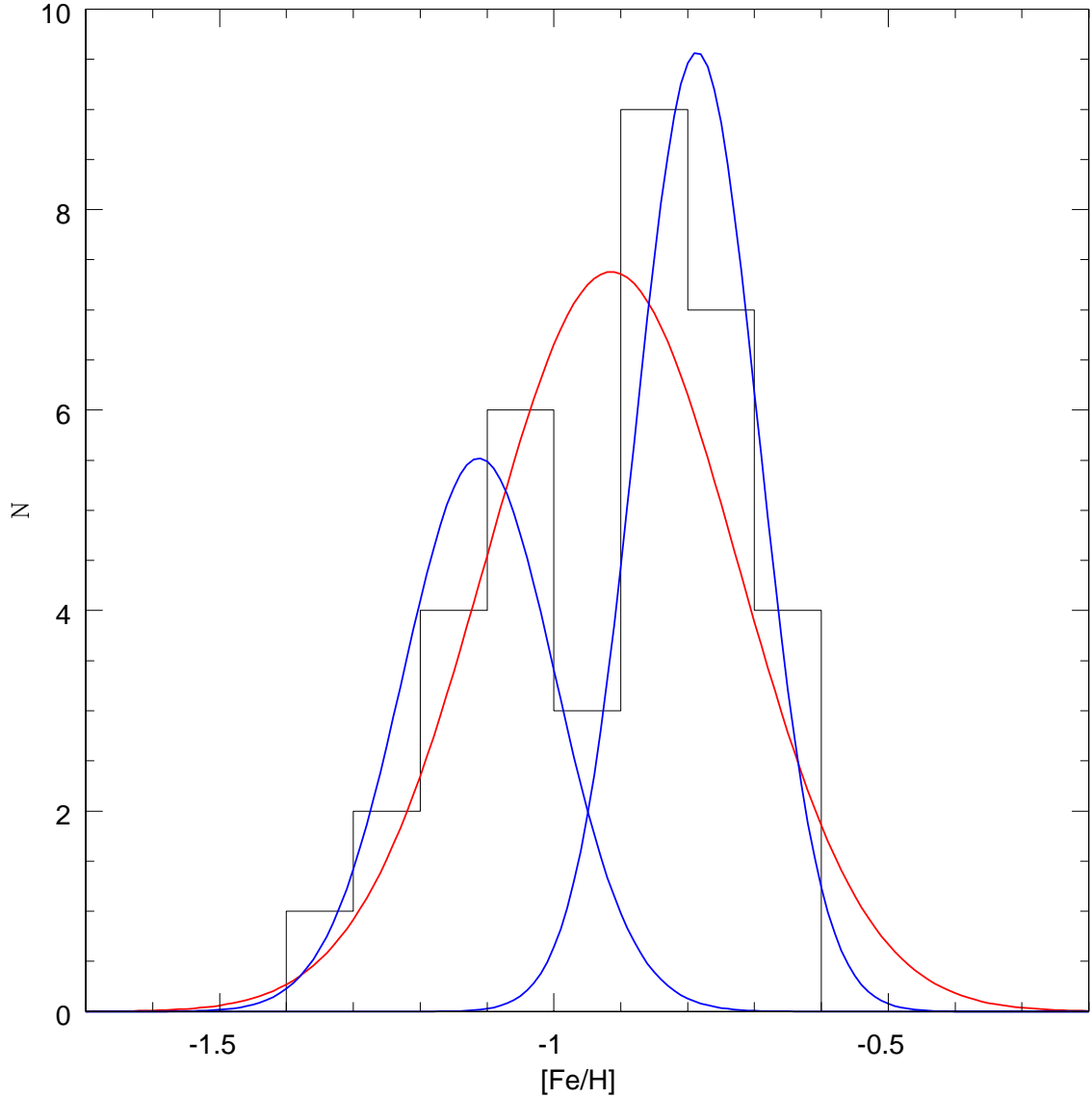}
\caption{Metallicity distribution of SMC clusters: 14 from the present work, 15 from P09,
3 from DH98, 3 from \citet{gla08b} and NGC\,330 \citep{gon99}.
Red and blue lines represent the unimodal and bimodal fits, respectively, according to the GMM algorithm independent of the bin selected for 
plotting this histogram. }
\label{f:MD}
\end{figure}

Figure \ref{f:MD} shows the MD of the 36  SMC clusters. It can be seen that this distribution suggests the existence of bimodality
with possible peaks at about [Fe/H] = -1.1 and -0.8, respectively. For a more quantitative analysis of the possible existence of bimodality, we applied the GMM (Gaussian Mixture Model, \citealt{mur10}). The unimodal fit (red line in Figure \ref{f:MD}) gives $\mu$ = -0.914 and $\sigma$ = 0.189, while the fit of two Gaussians (heteroscedastic split, blue lines in Figure \ref{f:MD}) gives $\mu_1$ = -1.112, $\mu_2$ = -0.786, $\sigma_1$ = 0.114 and $\sigma_2$ = 0.092.
In the homoscedastic case ($\sigma_1$ = $\sigma_2$), we found $\mu_1$ = -1.125, $\mu_2$ =
-0.793 and  $\sigma$ = 0.102. We obtained p values of 0.042 and 0.16 for the homoscedastic and heteroscedastic fits, respectively. 
This means that there is a probability of 4.2\% (homoscedastic case) and 16\% (heteroscedastic case) probability of being wrong in rejecting unimodality.
These values are in agreement with the probability given by the parametric bootstrap of 86\% that the distribution is indeed bimodal. The GMM 
algorithm calculated the separation of the peaks, finding D = 3.36 $\pm$ 0.73 and a kurtosis value of -0.852. To accept a bimodal distribution, 
values of D $>$ 2 \citep{ash94} and kurtosis $<$ 0 are required. The values derived for D and kurtosis support, therefore, the probability of bimodality. 
It should be made clear that these results do not depend on the metallicity bin, since the GMM is not applied to the histogram (e.g., the one shown in Figure 
\ref{f:MD} ) but to the metallicity values.
In P09 we had already found a suggestion of bimodality in the cluster MD, but now the evidence is substantially stronger.
\citet{muc14} published results based on a high resolution spectroscopic survey of ~200 SMC red giant field stars performed with FLAMES (VLT). He found
a main peak at $\sim$ -0.9/-1.0 dex, with a secondary peak at [Fe/H] $\sim$ -0.6, and suggested bimodality was 
present. Note that his peaks are offset from ours by about 0.2 dex. Also, we find more objects in the
metal-rich peak instead of the metal-poor peak, although the difference is much less dramatic than found by \citet{muc14}. However, recently \citet{dob14} did not
find evidence of any secondary peak in a considerably larger field sample (more than 3000 stars), in agreement with our field sample (P10; Parisi et al. in preparation). 
Obviously, more clusters studied with reliable metallicities are needed to definitively investigate the bimodality of the SMC cluster MD.

\subsection{Metallicity gradient}

Another important aspect to examine is the possible existence of a metallicity gradient in the SMC. The existence or not of a gradient
can be crucial for understanding the stellar formation and evolution processes in this galaxy. Important efforts to confirm or deny the existence of a 
metallicity gradient in the SMC have been made recently (e.g., \citealt{pia07a,pia07b,car08,cio09}; P09; P10). Despite 
their efforts, these authors have not come to an agreement about the nature of any metallicity gradient. 
Quite recently, however, \citet{dob14} found a clear -0.075 $\pm$ 0.011 dex deg$^{-1}$ gradient, based on the spectroscopic CaT metallicities of about 3000 SMC red field giants. They found this change of metallicity with radius within 4$^{\circ}$ of the center, with no significant change beyond this. Previously, based on a smaller sample but also 
from CaT metallicities of field red giants, \citet{car08} found hints of a radial metallicity gradient.

In this study, we analize the possible existence of an SMC metallicity gradient using our sample of 36 star clusters. As in P09, our sample has 
been divided in two groups: (i) those clusters located at a distance from the SMC center less than 4$^{\circ}$ and (ii) those located beyond this value. 
This division was based on the suggestion of \citet{pia07a,pia07b} which sustains that the inner cluster mean metallicity ($a$ $<$ 4$^{\circ}$) is larger than the 
mean metallicity of the clusters located in the outer region ($a$ $>$ 4$^{\circ}$). A total of 25 clusters in our sample lie in the SMC inner region (within 4$^{\circ}$), 
while 11 of them lie in the outer region. The mean metallicities (and their respective standard deviations) are -0.88 (0.18) and -1.00 (0.19) 
for the SMC inner and outer regions, respectively. The associated standard error of the mean are 0.04 and 0.06 dex for inner and outer clusters, respectively, which implies a statistical 
significance of $\sim$ 1.7 sigma between the two mean values. 

\begin{figure}
\plotone{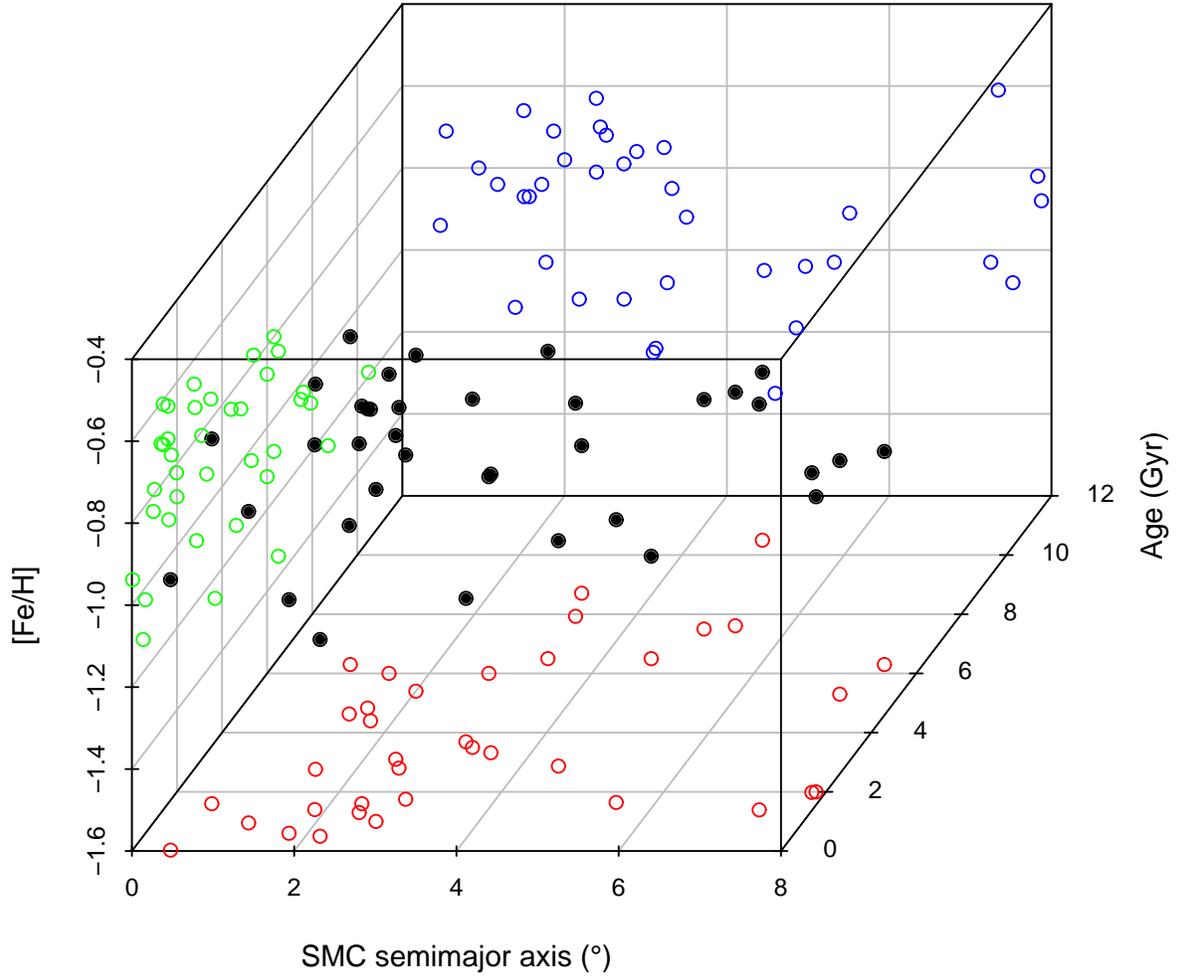}
\caption{ A 3D plot using the quantities [Fe/H], age and semi-major axis $a$ for our 36 cluster sample. The projection on each plane is represented in different colors.}
\label{f:3D}
\end{figure}

Figure \ref{f:3D} shows a three-dimensional representation of our cluster sample, the dimensions being metallicity ([Fe/H]), age and semi-major axis $a$. The 
age-metallicity relation (AMR) will be analized in the next section. Here, we focus on the projections on the (age,$a$) and ([Fe/H],$a$) planes. If only these 
two projections were considered, there seems to be a general trend of the metallicity to decrease with distance from the galaxy center $-$ at least within 
the first 4$^{\circ}$ from the galaxy center $-$ and of the age to increase with distance. However, it is worth considering 
that both the age and metallicity dispersions are remarkably large. Note the curious V-shape presented by the metallicity distribution in the ([Fe/H],$a$) diagram 
with the vertex around 4-5$^{\circ}$, which is also visible in \citet{pia11c}.

\begin{figure}
\plotone{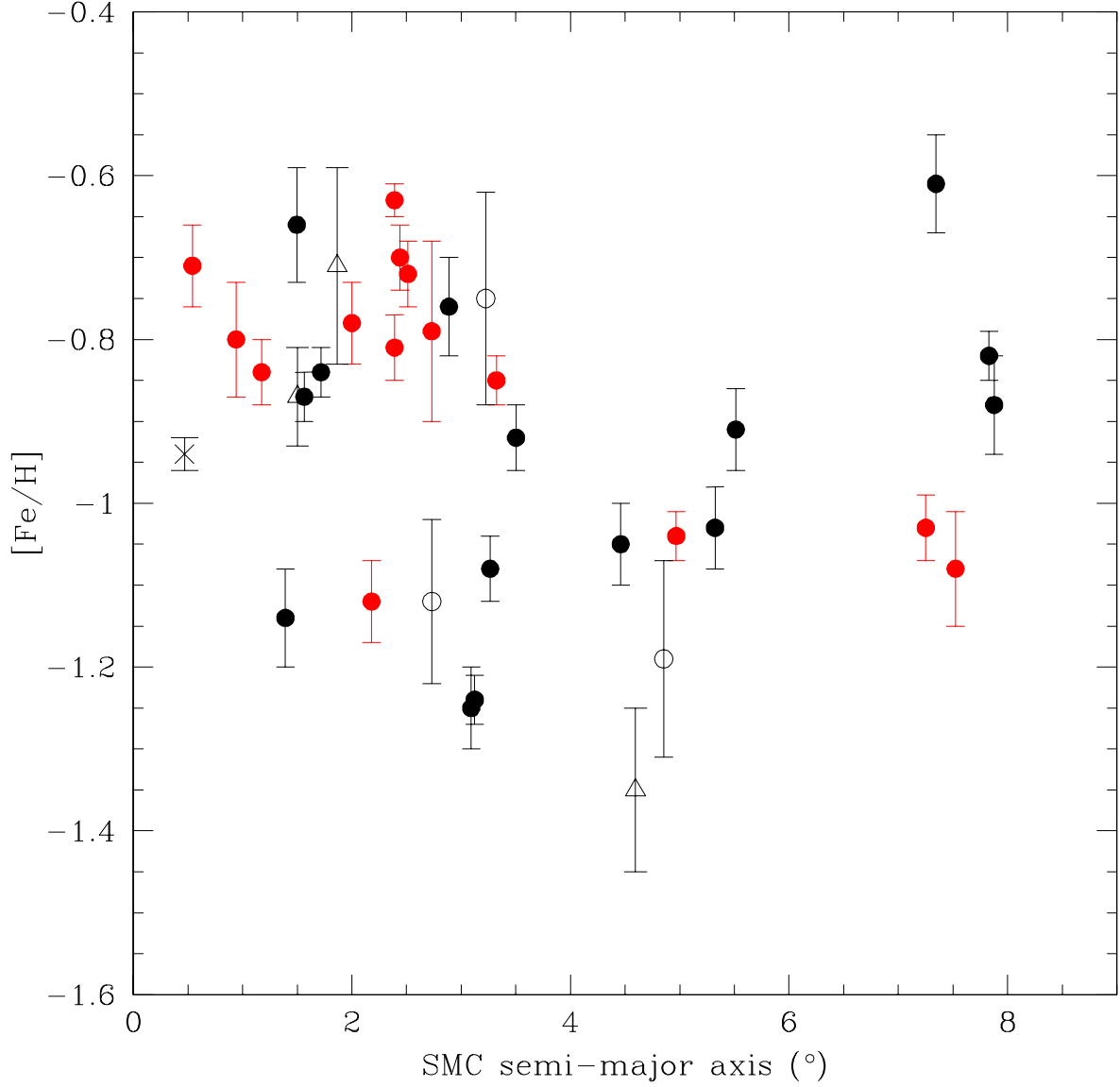}
\caption{Cluster [Fe/H] as a function of semi-major axis $a$ for the sample of 36 SMC clusters described in Figure \ref{f:MD}. Red and black circles represent clusters from the present work and P09, respectively. Clusters from DH98 are shown by open circles while triangles are clusters from \citet{gla08b}. NGC\,330 is represented by a cross.}
\label{f:met_axis}
\end{figure}

The behavior of the metallicity as a function of the semi-major axis $a$ for the 36 star clusters of our sample can be observed in Figure \ref{f:met_axis}. The meaning 
of the different symbols is explained in the figure caption. Although this figure suggests the possibility that the metallicity decreases with the semi-major axis $a$ (at least within the first 4-5$^{\circ}$ from the galaxy center), in agreement with \citet{dob14} findings, it is difficult to assert 
that there is a metallicity gradient in our cluster sample. This is mainly due to the large metallicity dispersion for each $a$ value, which may be as large as 0.5 dex. 
The weighted and unweighted linear fits of the data within 4$^{\circ}$ give slope values of -0.04 $\pm$ 0.04 and -0.05 $\pm$ 0.04, respectively. Both the slopes and 
their errors 
have comparable values; thus the fits are not statistically significant,  although the formal value we find (-0.05 $\pm$ 0.04) is in
reasonable agreement with that of \citet{dob14} (-0.075 $\pm$ 0.011). As for ages, it is necessary to remember that the cluster ages here 
analyzed are on a less homogeneous scale than the metallicities. \citet{dob14} interpreted the metallicity gradient due to a larger fraction of young 
stars toward the center of the galaxy, in concordance with other authors in previous works (e.g., \citealt{car08,wei13,cig13}). Considering the ages of our 36 star clusters, the ratio of clusters younger than 4 Gyr to older changes from 1.5 inside 4$^{\circ}$ to 0.83 outside. 
These numbers  are in agreement with \citet{dob14}'s idea but it is hard to assess the statistical significance given the small sample. Also in P14 we find no age gradient from a sample of 50 star clusters with ages determined in
a similar scale. Therefore, although we believe there is a suggestion of a metallicity gradient in our cluster sample, at least within the innermost 4-5$^{\circ}$, 
we cannot confirm it. Beyond that, any potential gradient becomes very flat or in fact turns around and rises. Note that indeed this is the case in the Galaxy, where 
the disk metallicity gradient flattens out beyond about 10-12 kpc (e.g., \citealt{twa97}). It is worth 
mentioning that a sample of $\sim$ 750 red field giants (P10; Parisi et al. in preparation) shows a metallicity gradient in the inner 4$^{\circ}$  in reasonable 
agreement with that of \citet{dob14}, with the metallicity dispersion of field stars much lower than that of the clusters. \citet{dob14} found that the 
SMC fields studied in P10 clearly exhibit a metallicity gradient, at least within the first 4$^{\circ}$ from the galaxy center. The problem arises when clusters are studied
alone or combined with field stars, in which case the large metallicity dispersion of the clusters blurs the gradient. 

\begin{figure}
\plotone{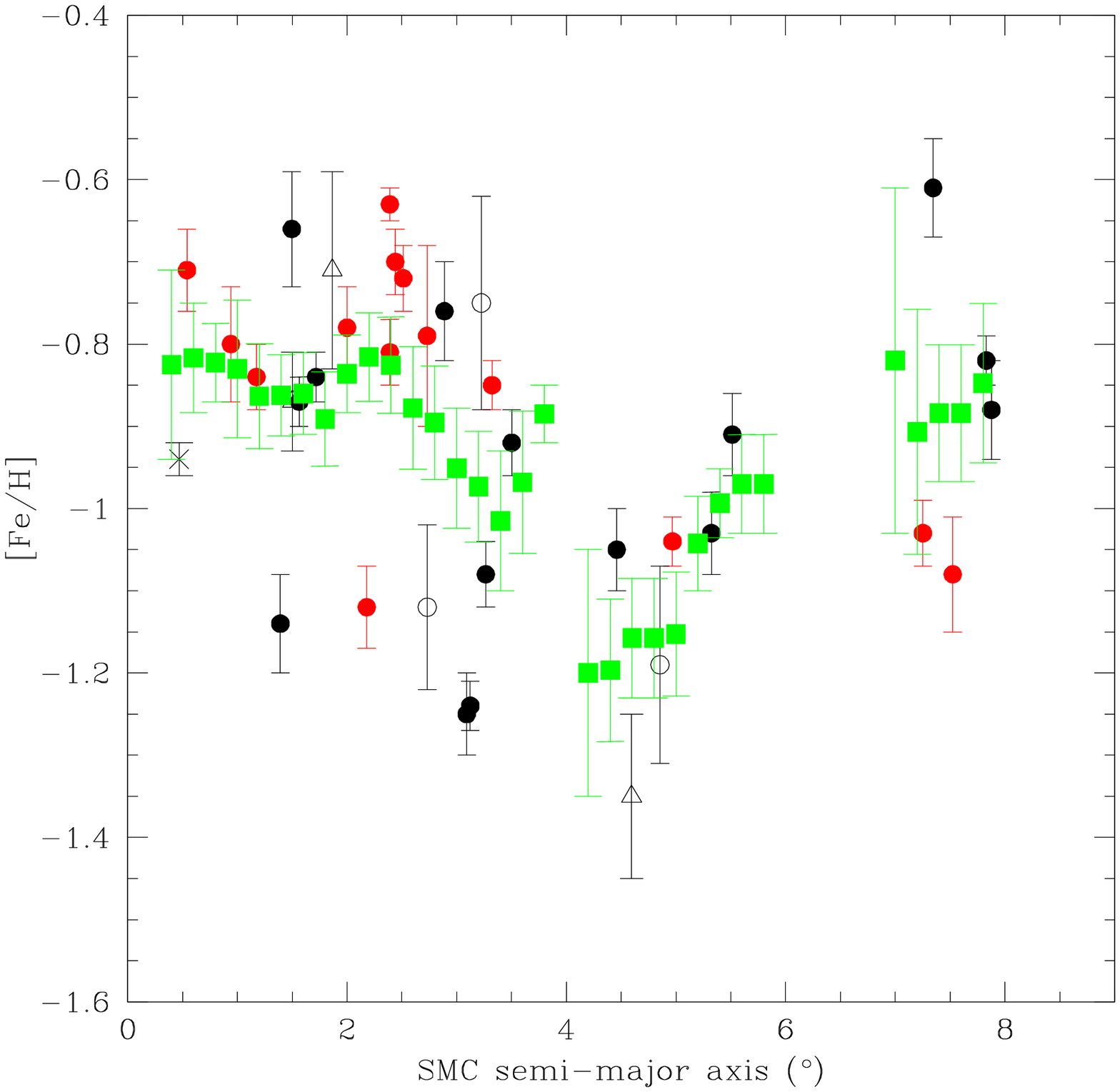}
\caption{Mean metallicity vs. mean semi-major axis $a$ (green squares). Error bars correspond to the standard error of the mean. The other symbols are the same as in Figure \ref{f:met_axis}. }
\label{f:v}
\end{figure}

To further investigate the above mentioned V-shape in the ([Fe/H],$a$) diagram, we divided the $a$ parameter in steps of 0.2$^{\circ}$. We then chose those clusters 
within $\pm$0.5$^{\circ}$ in each bin and calculated the mean metallicity and the error of the mean (green squares in Figure \ref{f:v}). The remaining symbols in 
Figure \ref{f:v} have the same meaning as in Figure \ref{f:met_axis}. We note that from 0$^{\circ}$ to 2.5$^{\circ}$, the trend of the metallicity appears to be flat. Then, the 
mean metallicity decreases exhibiting a minimum at $\sim$ 4-5$^{\circ}$. Finally, the mean metallicity rises to become flat again in the outermost parts. 
If this behavior is real, it is indeed curious and difficult to explain, especially if the large metallicity dispersion is taken into account. 

\subsection{Age-Metallicity relation}

\begin{figure}
\plotone{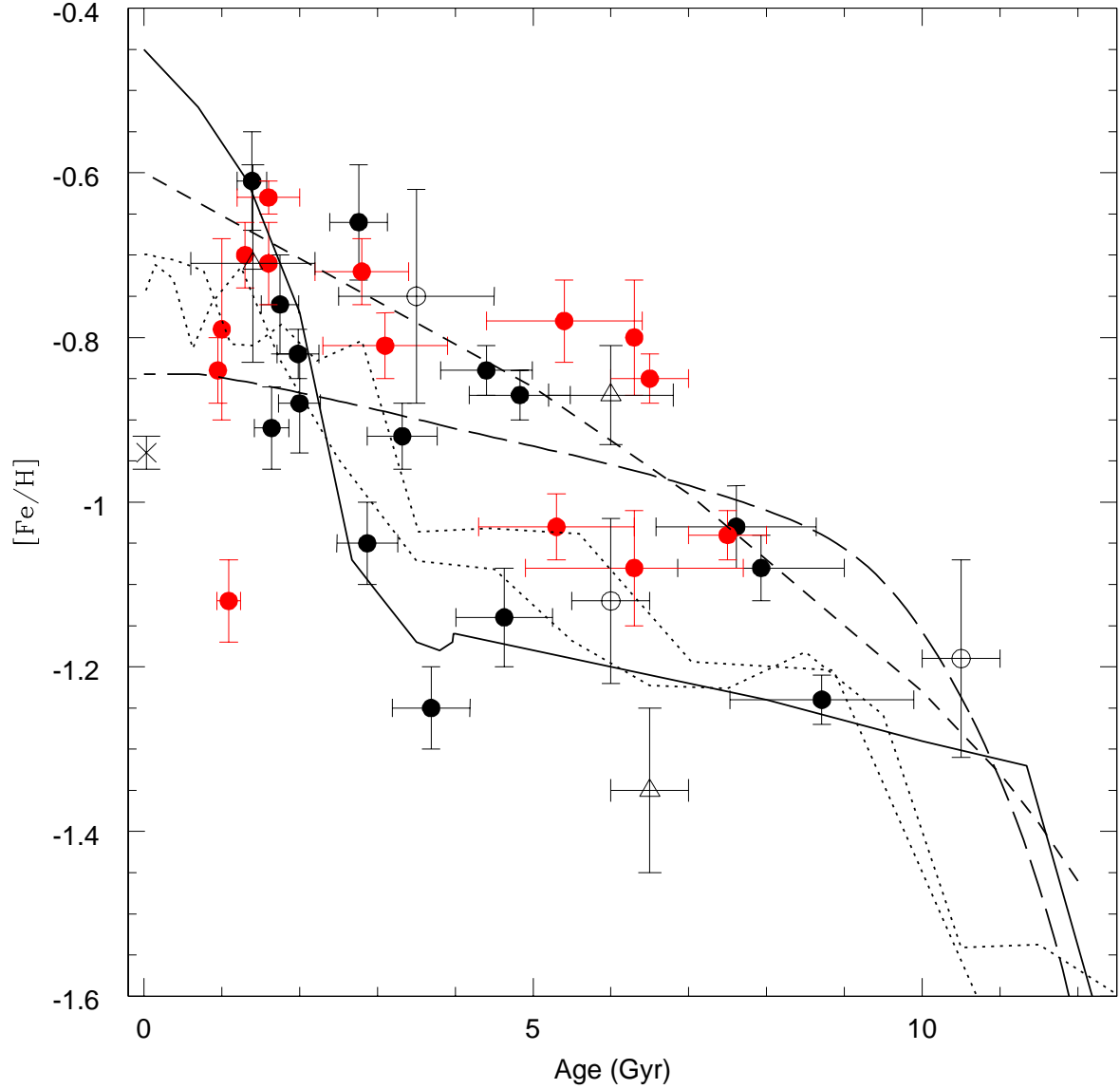}
\caption{Age-Metallicity Relation for the full sample (see caption of Figure \ref{f:met_axis} for details about symbols). Observations are compared with different models:
DH98 (short dashed line), PT98 (solid line), \citet[long dashed line]{car05} and \citet[dotted lines]{cig13}.}
\label{f:amr}
\end{figure}

Figure \ref{f:amr} shows the relation between age and metallicity for the 36 clusters of our sample. Symbols in this figure hold the same meaning as in Figure \ref{f:met_axis}. In an
attempt to understand how the SMC chemical evolution occurred, the observational data were compared with different models currently available in the literature.
The solid line in Figure \ref{f:amr} represents the PT98  bursting model, which posits star formation bursts.
Such an initial burst could have been followed by a long period with no chemical enrichment whatsoever between 11 and 4 Gyr ago, while a more recent star formation burst
could have considerably increased metallicity in the SMC. The short dashed line corresponds to the simple closed box model (DH98),
which assumes that the SMC chemical enrichment took place continuously and gradually throughout the galaxy's lifetime. The long dashed line represents the best fit
found by \citet{car05} for a large field star sample studied using the CaT technique. Finally, the dotted lines are the age-metallicity relations (AMRs)
obtained by \citet{cig13} from the study of six SMC fields. \\

In general terms, it is observed that for each age interval there exists a metallicity dispersion of $\sim$ 0.5 dex, which is very significantly larger than the corresponding
metallicity determination errors. Clusters do not appear to favor any of the models currently available. This seems to suggest that there is not a unique cluster AMR in the SMC. Either the galaxy was not well mixed chemically initially and/or the chemical evolution was
not a simple, smooth, global process, with the clusters being more
affected by dispersion processes than their offspring, the field stars.

\begin{figure}
\plotone{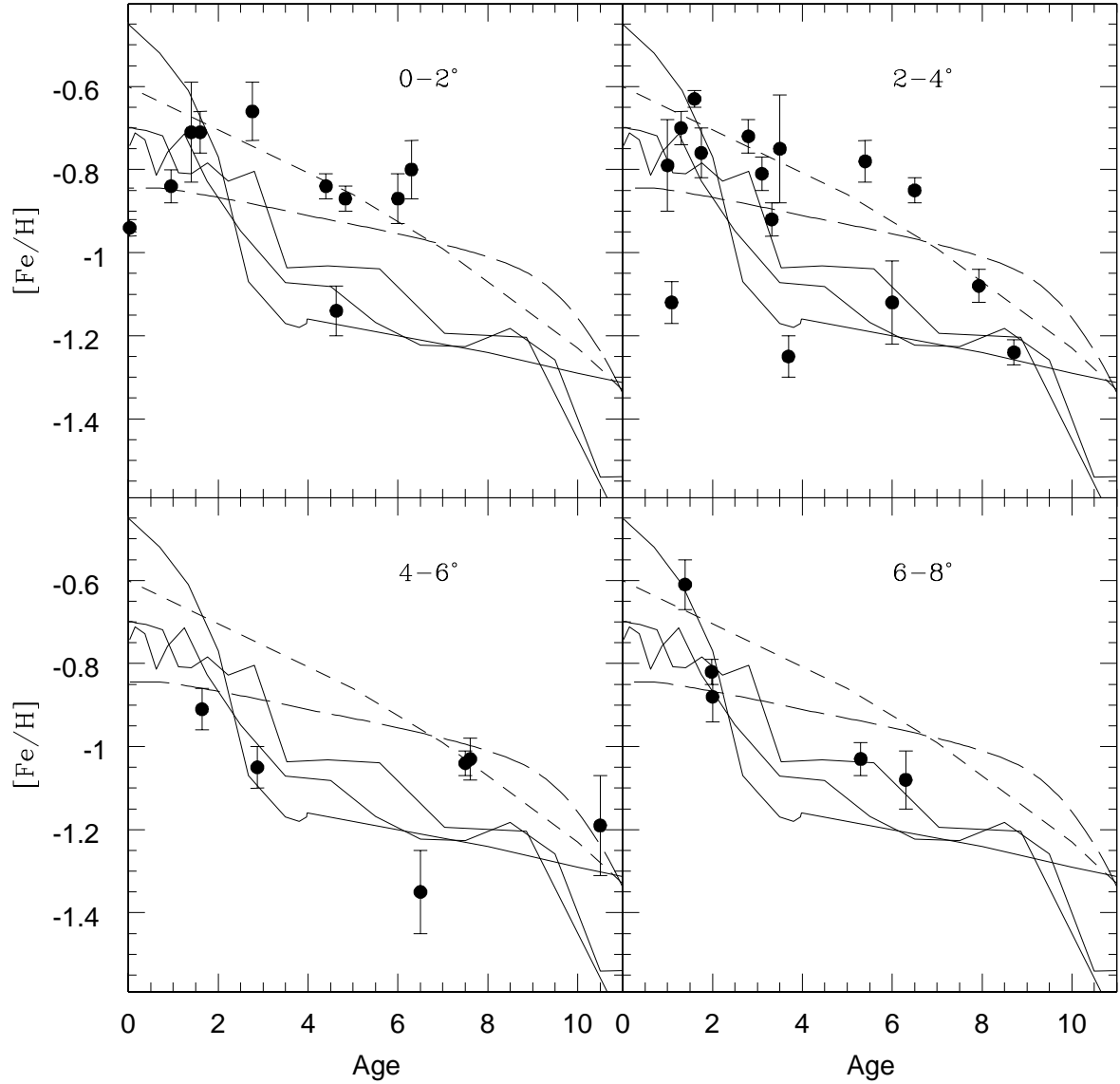}
\caption{Age-Metallicity Relation at different distances from the galaxy center. The
corresponding intervals in semi-major axis $a$ are shown in the panels. Lines are the
same that in Figure \ref{f:amr}.}
\label{f:amr_axis}
\end{figure}

Taking these findings into account, it would be
interesting to analyze the behavior of cluster metallicities in relation to their ages in different regions of the SMC.
Our cluster sample is still statistically too small to examine the AMR in particular regions of the SMC. However, a first approach to this study can be carried
out by examining the AMR at different distances from the SMC center. Our sample was divided in four groups according to semi-major axis $a$,
namely: 0$^{\circ} - $  2$^{\circ}$, 2$^{\circ} - $ 4$^{\circ}$, 4$^{\circ} - $ 6$^{\circ}$ and 6$^{\circ} - $ 8$^{\circ}$. The resulting AMR for each of these
four groups can be observed in Figure \ref{f:amr_axis}. The corresponding $a$ intervals are shown in each panel. The figure includes the same models as Figure \ref{f:amr}. Note that
clusters situated at distances from the SMC center larger than 4$^{\circ}$ seem to reasonably follow the bursting model. The same does not occur with the
inner clusters that better fit the tendencies of DH98 and \citet{car05} models. We point out, however, that in the outermost two $a$ intervals, the number
of clusters is not significant enough to achieve conclusive results. On the other hand, there are at least 6 clusters in the first two $a$ intervals considerably
more metal-poor than predicted by \citet{car05} and DH98 models. It is then absolutely necessary to have a larger cluster sample available so as to
examine any possible AMR variations in greater detail.

\begin{figure}
\plotone{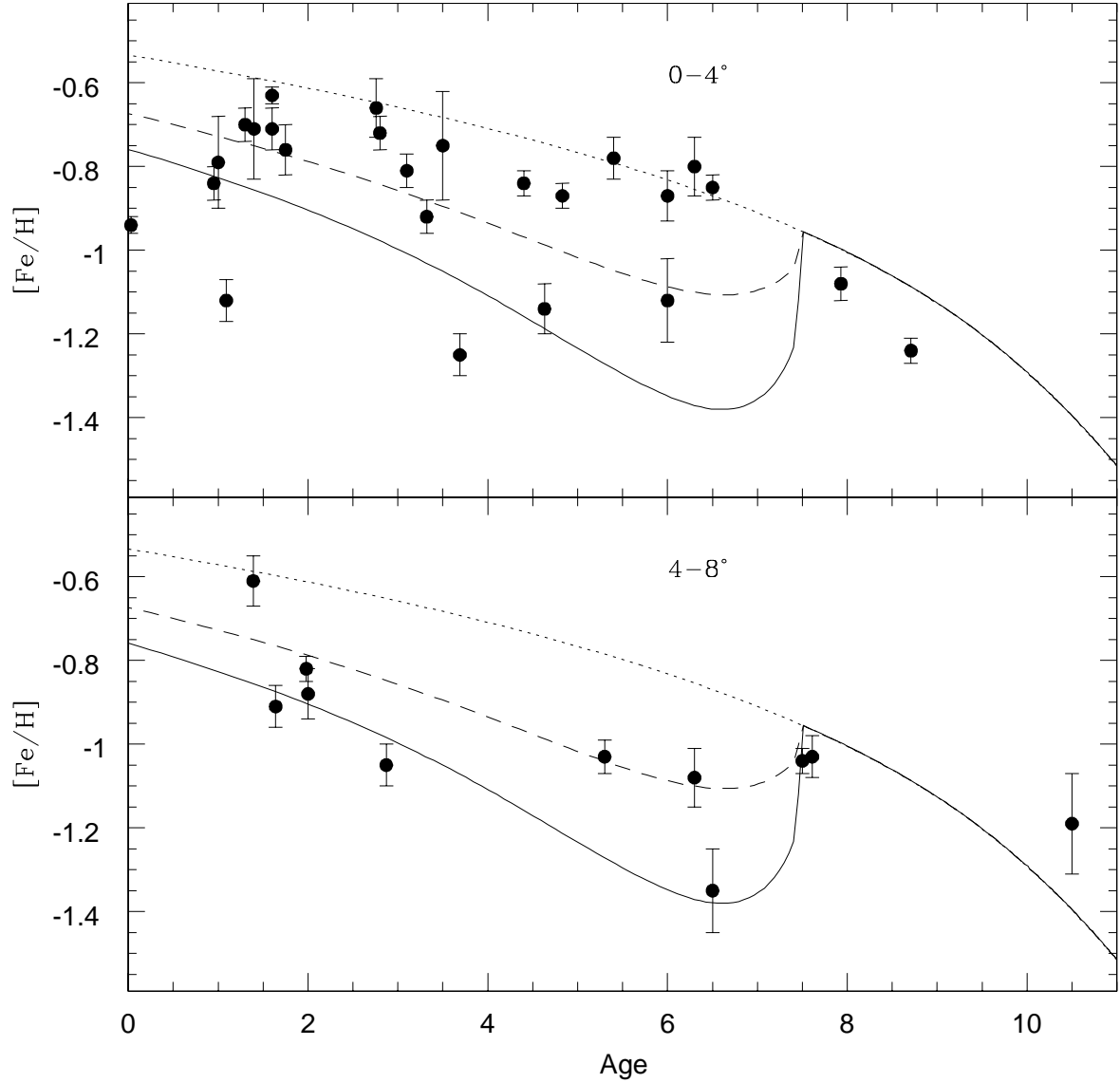}
\caption{Age-Metallicity Relation in two semi-major axis $a$ intervals, shown in the panels. Lines are the models from \citet{tsu09}. Dashed and solid lines represent
merger models with mass radio of 1:1 and 1:4, respectively. Dotted line is the model with no merger. 
}
\label{f:amr_TB}
\end{figure}

Several years ago, \cite{tsu09} computed two models of chemical evolution considering the merger of two galaxies with different mass ratios (1:1 and 1:4). 
They also computed a model with no merger. They compared these models with the AMR of cluster and field stars taken from the literature, derived from photometric \citep{pia01,pia05,pia07b,pia07c,gla08a,gla08b} and spectroscopic \citep{dch98,kay06,car08} studied. They suggested that a major 
merger occurred in the SMC $\sim$ 7.5 Gyr ago, evidenced by a dip in the AMR around that time. The cluster sample used by \cite{tsu09} is very heterogeneous in both age 
and metallicity. Also, the photometric metallicities are considerabley less precise that the spectroscopic ones (P09). With this in mind, it is of interest to compare the three models from \cite{tsu09} with the sample here studied, which is homogeneous, especially in metallicity values. Figure \ref{f:amr_TB} shows our 
AMR in two intervals of $a$ (0$^{\circ} - $ 4$^{\circ}$ and 4$^{\circ} - $ 8$^{\circ}$). Solid and dashed lines represent models of mergers having a mass ratio 
of 1:4 and 1:1 respectively, while the dotted line represents the model without a merger. 

It can be seen in Figure \ref{f:amr_TB} that when distances from the galaxy center are smaller than 4$^{\circ}$, our observations do not seem to favor any of 
the scenarios proposed by \cite{tsu09}. The large metallicity dispersion is the salient feature. In this $a$ interval, the metallicity 
dip proposed by \cite{tsu09} is not as clearly as in these  authors' work.  Conversely, our observations distinctly tend to favor merger scenarios in 
those clusters located beyond 4$^{\circ}$. The two models that assume mergers reproduce the clusters within this interval fairly well. There is, however, an 
exception: the cluster located at ($\sim$ 1.5 Gyr, $\sim$ -0.6 dex). Our evidence thus suggests the possibility that the SMC had suffered a merger event affecting mainly the 
outer part of the galaxy, which would be a possible explanation for the differences found between the SMC  outer and inner regions, mainly with regard to the metallicity gradient and AMR.

\section{Summary}

We used the FORS2 instrument on VLT to obtain spectra in the region of the CaT lines of more than $\sim$ 450 red giant stars belonging to 15 clusters in the SMC and their
surrounding fields. Following exactly the same procedure as in P09, we determined their metallicities and RVs with typical errors of 0.16 dex
and 7.3 km s$^{-1}$ per star, respectively, after discarding one cluster too young for the metallicity calibration. We analyzed cluster membership using as criteria distance from the cluster center, RV, and metallicity (G06, P09).
From those stars considered to be cluster members, we derived the
mean metallicity and RV of the 14 remaining clusters. We obtained mean cluster velocities and metallicities with mean errors of  2.6 km s$^{-1}$ and 0.05 dex, from a mean
of 6.5 members per cluster. Ages of our 14 clusters were adopted from the literature. \\

Using this information, together with that for other clusters similarly studied, we analyze the chemical properties of the SMC cluster system.
Specifically, we added to our present sample 15 clusters whose metallicities and ages were derived in P09 and P14, respectively, and also included 7 other clusters previously studied by DH98, \citet{gla08a,gla08b} and \citet{gon99}. The metallicities  of these
additional clusters  have been determined from CaT except one, NGC 330, whose metallicity was inferred from high dispersion spectroscopy.
Consequently, we compiled a sample of 36 SMC clusters with accurate metallicities. So far this is the largest SMC cluster sample with accurate and homogeneous metallicities. \\

Our main results are the following:

\begin{itemize}

\item The MD of our cluster sample appears to be bimodal, with potential peaks at $\sim$ -1.1 and -0.8 dex. We applied the Gaussian Mixture Model
\citep{mur10}, obtaining a high probability  that our data can be represented by two as opposed to a single Gaussian function.

\item Our data show a tendency of metallicity to decrease with distance from the center of the galaxy, at least out to about 5$^{\circ}$,
where any potential gradient appears to
flatten or even turn around, but we can not confirm the existence of a metallicity
gradient in our cluster sample, mainly because of the large metallicity dispersion.

\item We corroborate the P09 finding, now with a larger sample of clusters, that the AMR presents a significant metallicity dispersion of about 0.5 dex,
a value that exceeds the errors associated with the determination of the metallicity. This dispersion of metallicities may be evidence that there is not a
 single AMR in the SMC. In fact, none of the chemical evolution models currently available in the literature turn out to be a good representation of the general trend of [Fe/H] with age.
A larger statistical cluster sample is needed to analyze the AMR in different regions of the SMC.

\end{itemize}

\acknowledgments

This work is based on observations collected at the European Southern Observatory,
Chile, under programs number 082.B-0505 and 384.B-0687. We would like to thank the referee for the very complete and detailed comments which helped to
improve the manuscript. We would like also to thank the
Paranal Science Operations Staff. M.C.P. acknowledges Dr. Michele Cignoni for providing us
his age-metallicity relations and for interesting discussions, Dr. Tsujimoto for kindly providing us with his models, 
and Dr. Carlos Briozzo for his explanations 
on statistical concepts. 
M.C.P. and J.J.C. gratefully acknowledge financial support from the Argentinian
institutions CONICET, ANPCyT and SECYT (Universidad Nacional de C\'ordoba). D.G. gratefully 
acknowledges support from the Chilean  BASAL   Centro de Excelencia en Astrof{\'\i}sica
y Tecnolog{\'\i}as Afines (CATA) grant PFB-06/2007. S.V. gratefully acknowledges the support 
provided by Fondecyt reg. n. 1130721.

{\it Facilities:} \facility{VLT: Antu (FORS2)}.

\begin{deluxetable}{lccccc}
\tablewidth{0pt}
\tablecaption{SMC Clusters sample}
\tablehead{
\colhead{Cluster}                & \colhead{RA (J2000.0)}  &
\colhead{Dec (J2000.0)}          & \colhead{$a$}           &
\colhead{Age}                    & \colhead{Age}          \\
                                 & \colhead{($h$ $m$ $s$)} &
\colhead{($^{\circ}$ $'$ $''$)}  & \colhead{($^{\circ}$)}       &
\colhead{(Gyr)}                    & \colhead{reference}
     }
\startdata

B\,99, OGLE-CL\,SMC\,122        &       01 00 30.52 &   -73 05 14.40 &  1.174 & 0.95 &  1\\
B\,113                          &       01 02 55.75 &   -73 20 18.60 &  1.770 & 0.53 &  2\\
H\,86-97, OGLE-CL\,SMC\,43      &       00 47 53.42 &   -73 13 14.10 &  0.540 & 1.60 &  1\\
HW\,40                          &       01 00 25.11 &   -71 17 43.80 &  2.000 & 5.40 &  3\\
HW\,67                          &       01 13 01.82 &   -70 57 47.10 &  2.513 & 2.80 &  4\\
K\,3, L\,8, ESO\,28-19          &       00 24 47.70 &   -72 47 00.01 &  3.322 & 6.50 &  5\\
K\,6, L\,9, ESO\,28-20          &       00 25 26.60 &   -74 04 29.70 &  2.390 & 1.60 &  6\\
K\,8, L\,12                     &       00 28 02.14 &   -73 18 13.60 &  2.440 & 1.30 &  1\\
K\,9, L\,13                     &       00 30 00.26 &   -73 22 40.70 &  2.180 & 0.50 &  2\\
K\,37, L\,58                    &       00 57 48.53 &   -74 19 31.60 &  2.730 & 1.00 &  1\\
K\,44, L\,68                    &       01 02 06.34 &   -73 55 22.70 &  2.391 & 3.10 &  7\\
L\,1, ESO\,28-8                 &       00 03 54.00 &   -73 28 18.00 &  4.968 & 7.50 &  5\\
L\,112                          &       01 36 01.00 &   -75 27 00.30 &  7.524 & 6.30 &  8\\
L\,113, ESO\,30-4               &       01 49 30.00 &   -73 43 40.00 &  7.252 & 5.30 &  9\\
OGLE-CL\,SMC\,133               &       01 02 31.86 &   -72 19 05.30 &  0.941 & 6.30 &  1\\
\enddata
\tablerefs{
(1) \citet{rz05}, (2) \citet{par15},
(3) \citet{pia11a}, (4) \citet{pia11c},
(5) \citet{gla08b}, (6) \citet{pia05},
(7) \citet{pia01}, (8) \citet{pia11b},
 (9) \citet{pia07b}}.
\label{t:sample}
\end{deluxetable}

\begin{deluxetable}{lccccccccc}
\rotate
\tablewidth{0pt}
\tablecaption{Measured values for member stars}
\tablehead{
\colhead{ID}                     &  \colhead{RV}                  &
\colhead{$\sigma_{RV}$}          & \colhead{$v-v_{HB}$}            &
\colhead{$\Sigma W$}             & \colhead{$\sigma_{\Sigma W}$}  &
\colhead{[Fe/H]}                 & \colhead{$\sigma_{[Fe/H]}$}   \\
                                 &  \colhead{(km s$^{-1}$)}       &
\colhead{(km s$^{-1}$)}          & \colhead{(mag)}               &
\colhead{(\AA)}                & \colhead{(\AA)}                  &
\colhead{(dex)}               & \colhead{(dex)}
     }
\startdata

B99-6 & 158.2 & 5.5 & -1.59 & 7.25 & 0.07 & -0.762 & 0.109 \\
B99-7 & 167.5 & 5.5 & -1.76 & 7.42 & 0.09 & -0.745 & 0.113\\
B99-10 & 148.9 & 5.4 & -2.08 & 7.26 & 0.05 & -0.886 & 0.106\\
B99-15 & 162.9 & 5.5 & -1.19 & 6.30 & 0.08 & -1.001 & 0.102\\
B99-17 & 161.5 & 5.5 & -1.42 & 6.95 & 0.07 & -0.824 & 0.107\\
B99-21 & 156.1 & 5.4 & -1.12 & 6.68 & 0.06 & -0.844 & 0.105\\

\enddata
\tablecomments{Table \ref{t:giants} is published in the
electronic edition in its entirety. }
\label{t:giants}
\end{deluxetable}

\begin{deluxetable}{lccccc}
\tablewidth{0pt}
\tablecaption{Derived SMC Cluster Properties}
\tablehead{
\colhead{Cluster}       & \colhead{n}                 &
\colhead{RV}            & \colhead{$\sigma_{RV}$}     &
\colhead{[Fe/H]}        & \colhead{$\sigma_{[Fe/H]}$} 
                         \\
                        &                             &
\colhead{(km s$^{-1}$)} & \colhead{(km s$^{-1}$)}     &
\colhead{(dex)}         & \colhead{(dex)}             
     }
\startdata
B\,99             &     6  &    159.2 & 2.6 &   -0.84 &         0.04    \\
H\,86-97          &     7  &    124.5 & 2.8 &   -0.71 &         0.05    \\
HW\,40            &     3  &    142.1 & 2.0 &   -0.78 &         0.05     \\
HW\,67            &     4  &    110.0 & 3.1 &   -0.72 &         0.04     \\
K\,3              &     10 &    135.1 & 0.7 &   -0.85 &         0.03    \\
K\,6              &     6  &    161.0 & 2.1 &   -0.63 &         0.02    \\
K\,8              &     3  &    208.0 & 1.3 &   -0.70 &         0.04    \\
K\,9              &     7  &    113.1 & 3.1 &   -1.12 &         0.05    \\
K\,37             &     3  &    124.6 & 9.3 &   -0.79 &         0.11    \\
K\,44             &     13 &    165.1 & 1.1 &   -0.81 &         0.04     \\
L\,1              &     14 &    145.3 & 1.6 &   -1.04 &         0.03    \\
L\,112            &     6  &    175.8 & 2.3 &   -1.08 &         0.07    \\
L\,113            &     7  &    178.6 & 2.4 &   -1.03 &         0.04     \\
OGLE\,133         &     5  &    149.0 & 3.2 &   -0.80 &         0.07    \\

\enddata
\label{t:results}
\end{deluxetable}

\end{document}